\documentclass[aps,pre,showpacs,showkeys,twocolumn,letterpaper,floatfix]{revtex4}
\usepackage{graphicx}
\usepackage{dcolumn}
\usepackage{bm}

\begin{document}


\title{Opportunity for Regulating the Collective Effect 
of Random Expansion with Manifestations of Finite Size Effects 
in a Moderate Number of Finite Systems}


\author{Wlodzimierz Kozlowski}
\email{wlodekak@ibb.waw.pl}
\affiliation{Institute of Biocybernetics \& Biomedical 
Engineering\\of The Polish Academy of Sciences, Trojdena 4, 02-109 Warsaw, 
Poland}


\date{\today}

\begin{abstract}
One reports computational study revealing a set of general requirements,  
fulfilling of which would allow employing changes in ambient conditions to 
regulate accomplishing the collective outcome of emerging active network 
patterns in an ensemble of a moderate number of finite discrete systems. The 
patterns within all these component systems emerge out of random expansion 
process governed by certain local rule. The systems modeled are of the same 
type but different in details, finite discrete spatial domains of the 
expansion within the systems are equivalent regular hexagonal arrays. The way 
in which elements of a component system function in the local information 
transmission allows dividing them into two classes. One class is represented 
by zero-dimensional entities coupled into pairs identified at the array sites 
being nearest neighbors. The pairs preserve their orientation in the space 
while experiencing conditional hopping to positions close by and transferring 
certain information portions. Messenger particles hopping to signal the pairs 
for the conditional jumping constitute the other class. Contribution from the 
hopping pairs results in finite size effects being specific feature of 
accomplishing the mean expected network pattern representing the collective 
outcome. It is shown how manifestations of the finite size effects allow using 
changes in parameters of the model ambient conditions of the ensemble 
evolution to regulate accomplishing the collective outcome representation.
\end{abstract}


\maketitle


\section{\label{sec.1}Introduction}
This study contributes to a search for methods of {\itshape regulating} 
the collective outcome of assembly processes taking place in an ensemble of a 
moderate number $M$ of finite complex systems, of the same type but different 
in details, evolving simultaneously and independently, under influence of 
stochastic factors within equivalent finite spatial domains and in the same 
ambient conditions. The finite discrete spatial domains are equivalent regular 
arrays in all the component systems. Information transmission processes within 
these domains result in manifestations represented by various patterns of 
expanding spatial regions specified with the same feature in all the systems 
at each stage of the process. One assumes that these regions specified within 
all the component systems affect simultaneously certain receiver system.
Moreover, differences between contributions, which distinct component systems 
have to the effect of their influence on the receiver system, are attributed 
only to differences between spatial arrangements of the regions specified 
within these component systems. Accordingly, spatial arrangements of these 
regions within all the systems correspond to the receiver system 
transformation that manifests itself as a {\itshape collective effect} (CE) at 
each stage of the process. Thus the receiver system recognizes collective 
outcome of evolution processes within all the component systems. One also 
assumes that all the component systems contribute to the collective effect in 
the equivalent conditions. For the reasons just noted, the CE can be 
represented by certain pattern of distinguished regions within a spatial 
domain being equivalent to the domain of a component system.
Changing certain values parameterizing the ambient conditions of evolution of 
all the component systems is the way in which regulation of the collective 
effect is under consideration here.
\par
In course of this study we have found that a prerequisite for following this 
way is presence of {\itshape finite-size effects} (we use abbreviation, F-SE) 
i.e. specific features of the CE accomplishment which result from realizing 
the local information transfer in the component systems of the evolving 
ensemble by one dimensional, finite-size elements able to change their 
positions in finite discrete space while transmitting certain signals between 
their ends and preserving their orientation in the space. The signal is 
considered as finite portion of information transmitted discretely between an 
element ends. Accordingly, each such element is represented by the pair of its 
ends only, one end that only sends a signal and the other receiving it. These 
ends coincide with sites of the regular array which are nearest neighbors. 
Hopping of the pair to a separate position close by requires obeying certain 
stochastic condition. Whole the pattern within each component system is being 
accomplished at each step of the evolution and can be composed only of sites 
being destinations for the receiving ends of the pairs that have experienced 
hopping at this step. This results in high diversity of the patterns which 
manifests itself in appearance of rugged islands in course of a component 
system evolution initiated from a single cluster; this specific 
diversification is a prerequisite for appearance of the F-SE. It has been also 
found that manifestations of the F-SE allow regulating the CE accomplishment 
effectively, in the way just mentioned, if elements constituting other class 
signal the pairs to perform the conditional hopping. This class encompasses 
elements represented by zero-dimensional entities transferring the information 
portions by hopping between closest neighbor sites. The way of modeling the 
local information transfer just mentioned is explained in Section ~\ref{sec.2} 
and algorithms of modeling the conditional information transfer are presented 
in Section ~\ref{sec.3}. This has been assumed as simplest model allowing the 
manifestations of effects being the subject investigated in course of this 
research (this is elucidated in Sections ~\ref{sec.4} and ~\ref{sec.6}).
\par
Remarkable manifestations of the {\bfseries \itshape finite-size effects, 
finiteness and discreteness} have appeared indispensable to allow 
regulating the CE accomplishment. The number, $M$ of the component systems 
affects usefulness of these manifestations for the regulation. The term 
{\itshape moderate} $M$ used hereafter means: $M$ is so large that the 
collective effect is accomplished reliably with certain specific features and 
$M$ is so small that all diverse specific manifestations attributed to the 
discreteness, finiteness and transfers resulting in the finite-size effects 
are remarkable and can be easily identified in the CE. The way of estimating 
the lower bound $M_0$ of the $M$ being moderate and role of other values 
parameterizing the model ambient conditions are presented in 
Section~\ref{sec.4}. Criterion of detecting the upper bound, $M_1$ to the 
value $M$ considered as being moderate 
and its role are explained in Sections~\ref{sec.6.2.3},~\ref{sec.6.3}).
\par
One of the motives for doing this research is the necessity in
predicting the opportunity for regulating the collective effect emerging out 
of evolution of an ensemble of systems of nanoelements that are assembling 
quasi two-dimensional functional networks over prepatterned surfaces 
~(for the purpose of modeling this process, the nanoelements can be 
represented by pairs of their ends only in the way alluded to previously).
This task may be considered as being relevant to low cost process of 
self-assembling nanostructured functional materials 
(this is the so called {\itshape bottom-up method}) that are expected to be 
applied as sensors, electric field emitters stimulating living tissues and 
as other active covers. Then, an electron would be usually the signal 
transmitted by the nanoelements and this explains the convention of 
identifying 
them as nanowires. The short mobile nanowires would differ from one
another in length and conductivity, and the conduction would influence binding 
other elements at their ends. A number of them would be tethered, either 
temporarily or permanently, by elements of the surface texture that would 
constrain freedom of their displacements and the differences between the 
nano-wires would contribute to the system randomness. One may suggest a number 
of candidates for the material components of such systems.
For example, properties of single-walled carbon nanotube conductors \cite {1}
(see also \url{http://cnanotech.com}) allow considering their short intervals 
as candidates for the short nanowires.
The short nanotube intervals can emit electric field from their ends, form 
stable colloidal suspension in water \cite {2} and can be adsorbed 
spontaneously to a gold surface \cite {3}.
Possibility of fabricating multivalent receptors on the gold nanoparticle 
scaffolds \cite{4} seems to open the opportunity for employing them to 
recognize elements self-assembling into the network.
Moreover, gold nanocrystals can self-assemble into ordered hexagonal texture 
of the surface \cite {5} and  electron transfer appears to be engaged into 
their interactions with fullerene particles \cite {6}. This reveals 
feasibility of a material system evolution being relevant to the model process 
reported here.
Let us note, however, that general character of the model assumptions made in 
this research may make the presented approach useful for studying other 
physical processes also.
\par
A computational study of the opportunity for regulating the collective effect 
emerging out of the ensemble of evolving complex systems can be carried on by 
simulating the model process proposed here.
We employ the known idea \cite {7,8,9,10} of considering signals being 
finite portions of information about certain generalized model feature 
transported in discrete way between sites  of a finite discrete space (the 
space is represented here by finite regular hexagonal lattice). Appearance of 
that model feature at a site is indicated by covering this site. Thus a 
pattern of covered sites within spatial domain of a system represents effect 
of the system evolution. 
\par
It is specific to this model that the scheme of covering the sites is assumed 
so as to represent effectively the underlying transport mechanisms resulting, 
eventually, in the finite-size effects. For this purpose we adapt our method 
of discrete displacements, within frame of which hopping pairs of 
zero-dimensional entities have been employed to simulate effects of 
information transmission relevant to turbulent transport and, then, agreement 
with available experimental results has been demonstrated \cite{9}.
Here, the collective effect is 
represented by the pattern being mean expected form of the patterns emerging
under equivalent conditions, simultaneously and independently (in parallel) in 
all systems constituting the ensemble.
\par
In this work, one investigates random expansion process (REP) as Markov
process of covering sites of finite regular hexagonal array. States of the REP 
at stages $T$ are finite random sets (FRS), $F(T)=\{A_i(T) , i=1,2,...,M\}$. 
The realizations $A_i(T)$ develop simultaneously and independently in the
same model ambient conditions. Accomplishing of the model collective effect is
represented by sequence of patterns $\int\!F(T)$, each of which is mean
expected form of all the $A_i(T)$ and computed as mean measure set (MMS) after
known algorithm \cite{11} briefly reported in the Appendix~\ref{ap}. 
\par
Characteristics of the MMS evolution pattern are presented in Section 
\ref{sec.5}. Specific, predictable changes in the MMS evolution patterns, 
which result from varying certain parameters of the model ambient conditions, 
have been recognized as features attributed to the finite-size effects 
modeling. They reveal a possible way of searching for a method of regulating 
the collective effect emerging out of a material system ensemble for which the 
process underlying evolution of the ensemble may be adequately represented by 
the REP. These results are presented and discussed in Section~\ref{sec.6}. 
%
\section {\label{sec.2}Method of modeling the local information transfer} 
\begin{figure*}
\includegraphics{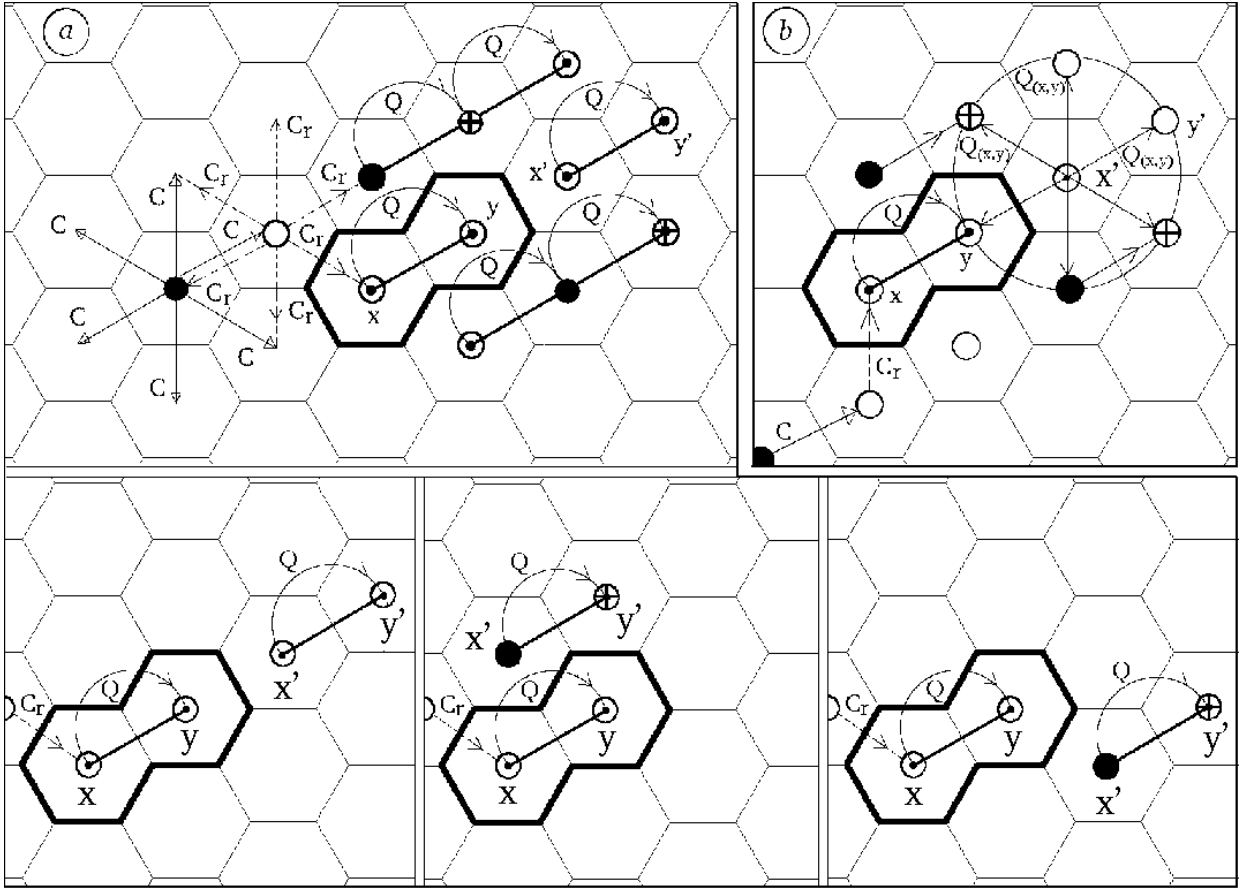}%
\caption{\label{fg.1} 
Effective modeling a path of the local information transfer in a component 
system $A_i$ for one step of the evolution, from $T$ to $T+1$.
(a) The nanowires represented by pairs of neighbor sites are activated by the 
messenger particle $C_r$  and experience hopping while being in state of 
transmitting a signal $Q$ (for the picture clarity, hopping from only one 
initial position of the nanowire is depicted). Three patterns at the bottom 
illustrate examples of the information transfer model events contributing to 
the virtual effective model event depicted above.
(b) Alternative representation of the local information transfer example: the 
$Q$ and the information about nanowire orientation, 
$(x,y)$ are considered as coupled into one transferable information portion, 
$Q_{(x,y)}$ ; note that activation of the nanowire does not depend on 
direction, from which the $C_r$ has come (see Section~\ref{sec.2}). 
In all the patterns: 
Black circles identify sites covered at stage $T$; the symbol, $\bigoplus$
detects sites that can be covered for the $T+1$ in result of the example event 
depicted; other sites which are indicated by the circles, $\bigcirc$
are not covered at the stage $T$ and will not be covered for the $T+1$
in result of the example event depicted (those uncovered sites, which coincide 
with ends of the nanowires shown, are indicated by $\bigodot$).
}
\end{figure*}

The way of modeling the mechanism of information transfer in each realization 
$A_i$ evolving from a stage $T$ to $T+1$ can be elucidated by 
considering certain idealized physical system. Let us assume that the 
system consists of short nanowires, transferring signals while hopping to 
positions close by, and of certain complex molecules occupying all sites of 
the finite hexagonal array representing texture of a prepatterned surface. 
Moreover, transmission of the signal changes binding properties at the 
nanowire ends and this can affect behavior of the complex molecules occupying 
sites coinciding with these ends. With the purpose to model effects produced 
by the nanowires, we generalize our approach \cite {9} used previously
for modeling the contribution which certain finite-size elements representing 
organized fluid streams and structures have to the collective effect of 
transport processes in the turbulent shear flow. In accordance with this 
approach, employed to situation being considered here, the finite-size 
elements transferring the signals are modeled as pairs of neighbor sites 
$(x,y)$ that transmit a signal from $x$ to $y$ while hopping effectively  to a 
position $(x',y')$ close by. This effective hoppings are realized so that 
information portion about orientation of two neighbor sites coupled into a 
pair is transfered from the position $(x,y)$ to $(x',y')$ (see the triplet of 
patterns in the bottom of Fig.~\ref{fg.1}a). Generally, each pair of 
neighbor sites can be considered coinciding with both ends of a model 
nanowire. Let us explain in this connection that the model being presented 
should not be considered literally: The model is effective in character , i.e. 
local model processes can represent cumulative outcome of all possible local 
actual realizations allowed in certain ambient conditions by the space 
configuration (here, hexagonal array) and related to one step of the model 
evolution. Further we explain spatiotemporal aspect of the effective modeling 
and thus reveal way of representing diversity of possible events which results 
from participation of the hopping pairs in the information transmission.
\par 
The pattern identifying the realization $A_i(T)$ is composed of sites 
covered at the stage $T$ only. 
The covering of a site represents activation of a complex molecule at the site
in the idealized physical system considered. In this system, each molecule 
activated sends identical messenger particles to all its nearest neighbors. 
In the computational model, each messenger particle is represented by a signal 
$C$ being the same for each site covered from which the $C$ is sent. 
Accordingly, each site occupied by such activated 
molecule is identified as a site with distributable feature $C$. One assumes 
that each molecule being active at a stage $T$ sends unconditionally the 
messenger particles $C$ to all its nearest neighbors at the step of 
evolution, from $T$ to $T+1$, and thus loses its activity. A molecule can 
become active at the stage $T+1$ only in result of a process identified 
hereafter as {\itshape transferring the molecule activity} by a nanowire at 
the step of evolution from $T$ to $T+1$; this process is elucidated below.
For first, we explain spatial aspect of the effective modeling of the local 
information transfer and this will help us to explain subsequently its 
temporal aspect.
\par
One assumes that the messenger particle $C$ coming to a molecule (either 
activated or not) from a nearest neighbor one can be effectively redirected to 
any molecule being the nearest neighbor of this intermediate one (see 
Fig.~\ref {fg.1}a). In virtue of the effective modeling alluded to previously, 
we represent contribution that these local redirection events have to 
effects of the transport process diversity as local cumulative result of the 
all possible model events of the redirection. Note that differences between 
the model events contributing to the local cumulative result concern only ways 
of the component redirections and are attributed to the space topology 
identified as regular hexagonal tiling. As no direction in this regular array 
is distinguished, the local cumulative result can be assigned to a virtual 
situation in which an active molecule would send for six messenger particles 
to each its nearest neighbor molecule that would redirect them, for one, to 
each of its six nearest neighbor molecules. The redirection 
does not cause any changes of the molecule redirecting the particles 
but the messenger particles redirected acquire new feature allowing them 
activating each nanowire whose sending end coincides with a site occupied by 
a molecule being nearest neighbor of the one that has redirected the particle.
This new feature is considered as enhancement of the signal $C$ due to the 
redirection and we denote the signal enhanced as $C_r$. Note that arrangement 
of sites into the hexagonal array allows activating nanowires at the six 
initial positions by one messenger particle $C_r$ achieving the site 
coinciding with their overlapping sending ends.
Here, for clarity, we continue the explanation for example of possible 
behavior of a nanowire at one initial position (sufficient conditions to be 
obeyed for activating those nanowires as well as nature of the effective 
modeling the transport process that engages them are revealed hereafter). A 
nanowire activated by the signal $C_r$ delivered to its sending end can act as 
a carrier hopping to a position close by while transmitting certain signal $Q$ 
emitted from its sending end toward the receiving one ~(see Fig.~\ref{fg.1}a). 
The $C_r$ does not affect the complex molecule at the site coinciding with the 
sending end of the nanowire. Moreover, state of this molecule, which can be 
intact or activated, has no effect on activating the nanowire by the $C_r$.
The $Q$ itself is {\itshape not} a molecule activation signal, transmission 
of the $Q$ by the nanowire is required to change binding properties at the 
nanowire ends: then the sending end can recognize a molecule that was active 
at a stage $T$ and this allows reception of 
the nanowire at its receiving end by the intact molecule occupying the same 
site as this end at the stage $T+1$. This reception makes this molecule active 
at the stage $T+1$. The reception of a single active nanowire is sufficient to 
activate the molecule. The molecule that can be activated in the way just 
presented will be identified hereafter as {\itshape target molecule} 
occupying the {\itshape target site}.
In accordance with assumptions alluded to previously, an active nanowire can 
experience hopping only to a position identified by a pair of neighbor sites 
so orientated in space as the pair of neighbor sites representing the 
nanowire at its initial position. Moreover, one assumes that directions of the 
nanowire displacements agree with direction of transmitting the $Q$, from the 
nanowire sending end to the receiving one. Accordingly, there are five such 
positions close by which are identified by pairs of neighbor sites that do 
not coincide with any of the two sites identifying the initial 
position ~(see Fig.~\ref{fg.1}a). 
Let us recall that, within frame of the information transfer 
representation being considered, the state of transmitting the signal $Q$ and 
hopping of the active nanowire take place at one step of evolution which is 
elementary unit of time here. Therefore, change of binding properties at the 
active nanowire ends due to appearance of the nanowire in the state of 
transmitting the signal $Q$ concerns also the nanowire at its initial position.
The nanowire action following the change in binding properties of its ends can 
be considered as direct transfer of the molecule activity. Eventually, we 
consider an active nanowire that can transfer directly the molecule activity 
at any of the six positions identified as {\itshape destination positions}.
Thus it can contribute to the information transmission stream that is 
initiated by emitting the messenger particles, $C$ and occurs at each step of 
the evolution. 
\par
Let us explain in this connection the nature of the effective modeling of the 
local transfer realized by the nanowires. Our approach is like the one used in 
the case of the messenger particle redistribution elucidated here previously. 
Contribution that the local hoppings of the active nanowires have to 
effects of spatial diversity in the transfer processes for one step of the 
evolution, from the $T$ to $T+1$, is modeled as result of a virtual event in 
course of  which six nanowires are activated at one initial position, 
experience hopping and are able to transfer directly the molecule activity at 
the six destination positions respectively ~(see Fig.~\ref{fg.1}a). Let us 
note that the pattern just sketched for example of a nanowire activated at one 
initial position remains the same for each of six possible initial positions 
at which nanowires with coinciding sending ends are activated by the $C_r$ 
(one messenger particle $C_r$ is sufficient to activate nanowires at all the 
six initial positions). Certain ambient gradient can constrain number of the 
initial positions at which activation of the nanowire is allowed (we elucidate 
that opportunity hereafter).
\par
Temporal aspect of the effective modeling may be explained by using a notion 
of virtual multistate of a complex molecule that has been activated at a stage 
$T$. At a step of the effective model process, it is thought as appearing in 
virtual three-state: the state of becoming intact by emitting the messenger 
particle $C$, the state of being probed by sending end of a nanowire at its 
destination position (then, available information about molecule activity 
refers to the stage $T$ only), the state of remaining intact or becoming 
active for the future by accepting receiving end of the nanowire. Eventually, 
the molecule appears in the single state of being intact or activated at the 
stage $T+1$. Accordingly, participation of the messenger particles and the 
model nanowire in the local information transfer is related only to the 
virtual three-state of the complex molecule as alluded to previously (this 
refers also to virtual bi-state of the complex molecule being intact at the 
$T$ and thus unable to emit the particle $C$).
\par
The {\itshape effective} model pattern of the local information transfer 
realized by the nanowires can be alternatively represented by a formal scheme
in which information about nanowire orientation, $(x,y)$ in the discrete space 
and the signal $Q$ are coupled into one transferable information portion 
denoted as $Q_{(x,y)}$ ~(see Fig.~\ref{fg.1}b). Within frame of this formal 
scheme, activation of a nanowire by the $C_r$ results in appearance of the 
information portions, $Q_{(x,y)}$ and their distribution to the six target 
sites being the nearest neighbor sites of the $x'$ identified as a site 
nearest the pair $(x,y)$ in the direction from $x$ to $y$. Thus, this $x'$ may 
be imagined as a node redirecting the hopping information portions, 
$Q_{(x,y)}$ toward all its closest neighbors. This imaginary event will be 
identified hereafter as act of {\itshape hopping-distribution}. In accordance 
with role of the signal $Q$, which has been explained here previously, 
delivering the $Q_{(x,y)}$ in the way of the hopping-distribution to each 
respective target site allows covering it at a stage $T+1$ if its 
nearest neighbor site, so oriented in respect to it as $x$ is oriented in 
respect to $y$, has been covered at the stage $T$ of the evolution. The 
$Q_{(x,y)}$ will be identified hereafter as information portion about covering 
such target site at $T+1$ (we will also use the phrase a {\itshape signal 
about covering a target site} to identify the $Q_{(x,y)}$). The scheme of the 
hopping-distribution is used, as more convenient, for presenting algorithms of 
the conditional information transfer.
\par
Let us recall that we model evolution that is random process and the scenario 
presented is to be supplemented by introducing certain stochastic condition. 
Note, in this connection,  that one of the outcomes of redirecting the 
messenger particle $C$ is reflecting this particle acquiring the property 
$C_r$ back to the molecule sending it. One may consider a process allowing 
only this way of redirection. Then, only the nanowires whose sending ends 
coincide with the site occupied by the molecule sending the $C$ can be 
activated by the messenger particle reflected as the $C_r$. This scenario 
corresponds to the process that can be considered as particular case of the 
process presented here previously. 
In this reduced scenario, the hopping-distribution would be the only model 
mechanism of the information transfer (we call it, direct transfer 
process (DTP) whereas the process encompassing the complete 
distribution of the $C_r$ is identified as indirect transfer process (ITP)). 
Note, however, that transfer of the information portion about orientation of a 
pair of neighbor sites in the discrete space is specific feature of the model 
process in both the scenarios, ITP and DTP. For that reason, one assumes a 
stochastic condition of occurrence of the information transfer realized by the 
active nanowires: in terminology used in the formal model, this is condition 
of occurrence of the hopping-distribution. 
Moreover, assuming the agreement between directions of an active nanowire 
displacements and direction of transmitting the $Q$, from the nanowire 
sending end toward the receiving one, suggests expressing this condition as 
relationship between a value characterizing local, time independent effect of 
stochastic factors and certain ambient gradient of a determined forcing factor.
Accordingly, we assume a constant value parameterizing one 
determined ambient gradient of a forcing factor imposed onto all the ensemble 
systems in course of the evolution. Note that using a gradient of the forcing 
factor to express the stochastic condition will result in constraining a 
number of the initial positions at which activation of a nanowire is allowed.
Value of the gradient pertaining to expression of this condition can be 
assumed so that evolution results in stable state without further growth 
of the lattice area covered (see Section~\ref{sec.4.1},~\ref{sec.5}). 
\section {\label{sec.3}Algorithms of conditional local transfer} 
\subsection {\label{sec.3.1}Spatial domain} 
All the independent realizations, $A_i(T)$ evolve in the same spatial domain 
under influence of stochastic factors and with the same process  parameters 
characterizing the model ambient conditions. The spatial domain is a finite 
discrete space  $\chi=\{x_k, \; k=1,2,...,N\}$ constituted by $N<\infty$ nodes 
of a regular triangular grid embedded into a plane so as to fill in a square 
with the sides of $1 \times 1$. This grid has largest density of nodes among 
all possible grids with the same value of the smallest distance between 
the nodes \cite {12}. 
Each of the embedded nodes is situated at center of a regular hexagon of the
honeycomb array covering the square. A node as well as the hexagon assigned to 
it are identified also with the term "site"  (the terms "node" and 
"site" are used interchangeably here). A node $x$ and its vicinity $V_x$
composed of its nearest neighbors $y_1,y_2,...,y_6$ are 
shown in  Fig.~\ref{fg.2}a ~(the $V_x=\{y_j\;,\;j=1 \div 6\}$ together with 
the $x$ are denoted as neighborhood $S_x$). The $A_i(T)$ are represented by 
patterns composed only of sites covered at stage $T$ of the evolution. 
\subsection {\label{sec.3.2}Direct transfer process REP-DTP} 
We employ the scheme of the hopping-distribution (see Section~\ref{sec.2}) to 
present algorithm of {\itshape conditional} covering the sites.
Let a node $x$ shown in Fig.~\ref{fg.2}b be
covered in the realization $A_i$ at a stage $T$,  $x \in A_i(T)$ (symbol 
$\in$ is read "is an element of"). This site is considered active and sends 
signals $C$ that can be reflected only by the neighbor sites, as signals 
$C_r$, back to the $x$. Accordingly to the assumption put forward in
Section~\ref{sec.2}, a model nanowire represented by the pair of neighbor 
sites at its initial position,  $(x,y_j) \; {\rm with} \; y_j \in V_x$, can be
activated by the signal $C_r$. Index, $j$ identifies here orientation of the 
pair within the $S_x$ and pair of indices, $jl$, with $l=1,2,...,6$, is used 
to indicate all target sites, to which the information portions, $Q_{(x,y_j)}$ 
are distributed if certain {\itshape condition} referred to stochastic factors 
is satisfied. The numbers are assigned to the indices in accordance with 
assignment shown in Fig.~\ref{fg.2}a ~(for clarity, only one example 
of the model nanowire at the initial position $(x,y_{j=3})$ is depicted in 
Fig.~\ref{fg.2}b and reflecting of the $C$ is not shown). The denotation, 
$x_{l}'$ is used to indicate sites being nearest the $y_{jl}'$ and so oriented 
in respect to the $y_{jl}'$ as the site $x$ is oriented in respect to $y_j$. 
In terms used for describing scheme of the hopping pairs (see, 
Fig.~\ref{fg.1}a), the denotation, $(x_{l}',y_{jl}')$ is used to indicate 
destination positions of the pairs. The rule of indexing presented here 
implies the equivalences, $x_{l=6}' = x$ and $y_{j,l=6}' = y_{j}$ (see the 
example for $j=3$ in Fig.~\ref{fg.2}b). The $y_{jl}'$ can be covered for the  
$A_i(T+1)$ if there is its neighbor site $x_l'$ covered at the stage $T$, 
$x_l' \in \{A_i(T) \cap V_{y_{jl}'}\}$ (here symbol $\cap$ denotes taking 
common part of the sets).  
\par 
The stochastic condition of the hopping-distribution of the $Q_{(x,y_j)}$ 
from the $S_x$ is assumed as requirement of obeying the
inequality $P_x(y_j) \geq r_x(y_j,i)$. Here, $r_x(y_j,i)$ 
is an element $r_{x_k}(\xi_j,i)$ of the pseudo-random matrix $\|r\|$
for  $x=x_k$, where $k=1 \div N$, $\xi_j=y_j$ for all 
$j=1 \div 6$, $\xi_{j=7}=x_k$ and $i=1 \div M$. 
The matrix $\|r\|_{N \times 7 \times M}$ is computed
by using a pseudorandom number generator (RANMAR \cite {13} or, when noted,
RANLUX \cite {14,15}) and independently of the stage $T$ of the REP 
development (this is held also for other variants of the processes reported 
here). The $P_x(y_j)$ is the probability assigned to a pair at the position 
$(x,y_j)$. The hopping-distribution of the $Q_{(x,y_j)}$ which may result from 
activation of the pairs from positions $(x,y_j)$ within 
$S_x$ are independent events. 
Here, the value $\mathop {\sum_{j}}P_x(y_j)  \leq 1$ is equal 
to a probability $P_s$ of occurring the signal transfers due to 
hopping-distribution of the $Q_{(x,y_j)}$ from the $S_x$ and the transfer will 
not take place due to stochastic factors with the probability $(1-P_s)$.
\subsection {\label{sec.3.3}Indirect transfer process REP-ITP}  
This variant of the REP corresponds to a process underlying evolution of a 
material system in which information about changing system features at 
a site can be distributed locally to engage more finite size elements 
into the transfer process. 
\par
\begin{figure*}
\includegraphics{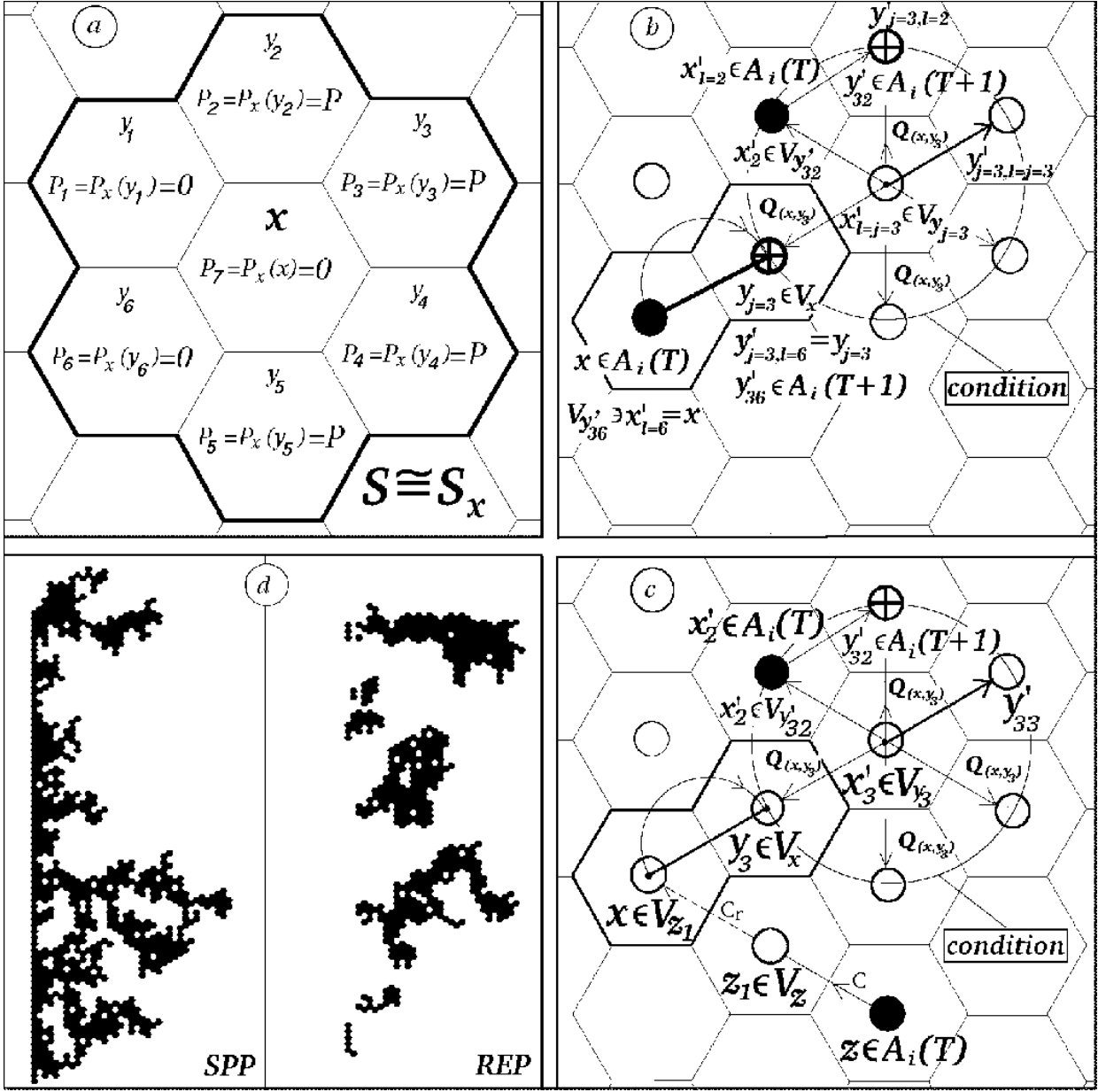}%
\caption{\label{fg.2} 
(a) A closest neighborhood $S_x$ of a site $x \in \chi$ with
assignment of the $P_j$.
Formal schemes of the local transfer, the example of one finite-size element
activated and act of the hopping-distribution is depicted:
(b) direct transfer process  REP-DTP; 
(c) indirect transfer process  REP-ITP; 
the box with the word, {\itshape condition} and arrow aiming at the big circle 
means that the hopping-distribution shown for example of the pair $(x,y_3)$ 
occurs if the condition, $P_x(y_3) \geq r_x(y_3,i)$ is satisfied; meaning of 
the circles at sites is explained in the caption to Fig.~\ref{fg.1}.
(d) Manifestation of a pattern diversity specific to the REP-ITP realization 
is revealed here by comparing this pattern 
to a realization pattern at a stage of the spreading percolation SPP 
(see also Section~\ref{sec.4}).
}
\end{figure*}

One assumes that the signals $C$ informing that a site $z$ is covered, 
$z \in A_i(T)$ are sent to all closest neighbors $z_1$ of the
$z$ ~($z_1 \in V_z$) and each $z_1$ redirects the signals. In this way, the 
$C$ is enhanced to $C_r$ that is delivered to all the $x \in V_{z_1}$ ~(see 
Fig.~\ref{fg.2}c). 
This communication from the $z$ to $x$ is effective if the stochastic 
condition of the hopping-distribution of the $Q_{(x,y_j)}$ is obeyed,  
$P_x(y_j) \geq r_x(y_j,i)$ and there is the site
$x_l' \in \{A_i(T) \cap V_{y_{jl}'}\}$  so oriented in respect to 
$y_{jl}'$ as the $x$ is oriented in respect to $y_j$ 
~(see Section~\ref{sec.3.2}). 
In this case, the signal, $Q_{(x,y_j)}$ is delivered to the $y_{jl}'$ that
is being covered, $y'_{jl} \in A_i(T+1)$. This scheme, identified as indirect 
transfer process (REP-ITP), allows exploring larger number of node 
configurations close to every site covered at a stage $T$ while searching for 
pairs realizing transfers resulting in covering target sites at 
the stage $T+1$. 
\par
Note that $x$ and $z_1$ can be or can not be elements of $A_i(T)$ 
and this ITP scheme allows also $x$ coinciding with the $z$ in the 
case of reflecting the signal by $z_1$. Thus local realization of the ITP
can include the DTP as particular case. Eventually, it may happen 
that $x \in A_i(T)$ when $x \neq z$ or $x \in A_i(T)$ because $x=z$. 
Both the cases correspond to contribution of the REP-DTP to  the REP-ITP.
Variant of the REP-ITP that allows only $x \notin A_i(T)$ will be
identified here as pure ITP ~(REP-ITP-P). 
\begin{figure*}
\includegraphics{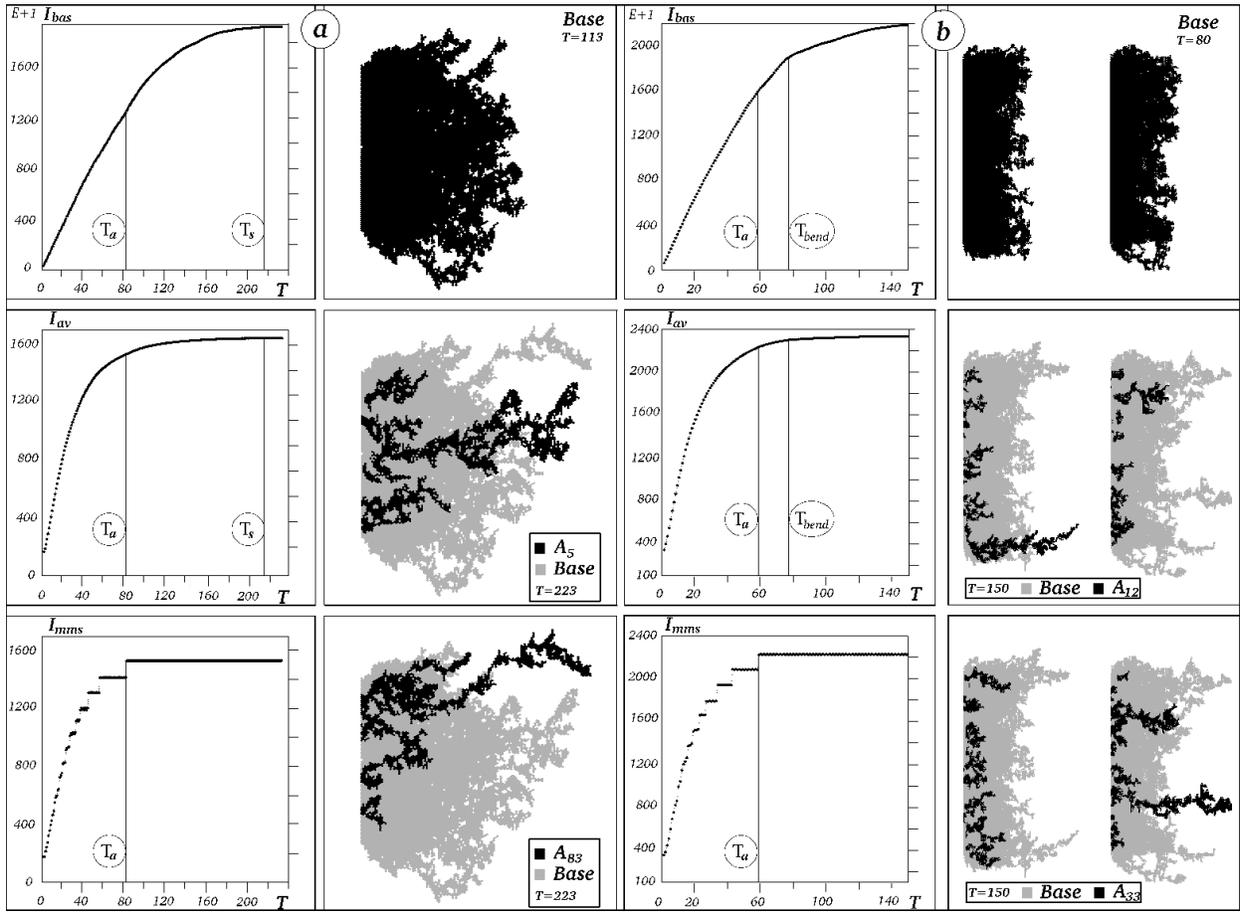}%
\caption{\label{fg.3} 
Characteristics of the REP advancement shown for examples of REP-ITP
representing two classes of evolution:
(a) without a bend in $I_{bas}(T)$, here,
{\large $\alpha$}$=0$, $M=M_0\!=\!100$, $P=0.24$ 
~(the sum $E+1$ means that the values shown should be multiplied by ten);
(b) with the bend in $I_{bas}(T)$, here,
{\large $\alpha$} $={1 \over 3}$, $M=M_0\!=\!100$, $P=0.1512$.
Two of six and two of five realizations developing yet at 
the very late stages are shown for examples of the classes (a) and (b), 
respectively.
The realizations are depicted as dark patterns imposed on the brighter pattern 
of the base.
}
\end{figure*}
\section {\label{sec.4}Parameters of ambient conditions}   
\subsection {\label{sec.4.1}Forcing parameter $P$} 
Ambient gradient of a forcing factor imposed onto all systems of the ensemble 
is one of the macroscopic conditions. Its effect is modeled in the same way in 
all the realizations $A_i$: 
by the same specific assignment of the probabilities $P_j$ to all pairs 
$(x,y_j)$ within the neighborhood $S_x$ for all the  $x \in \chi$ 
(see Fig.~\ref{fg.2}a$\div$c),
\begin {equation}   
P_1=P_6=P_7 \equiv 0 \:, \; P:=P_2=P_3=P_4=P_5 \; \leq 0.25 ,\label{eq.1}
\end {equation}
and, accordingly, by initiating the expansion from the same pair of
sufficiently long vertical chains (we identify them as the initiating 
structure IS; one considers two chains for the reliability purpose only,
see Fig.~\ref{fg.3}b).
The assignment of the $P_j$ is assumed the same for both the REP variants, 
ITP and DTP, being investigated here.
The value $P$ parameterizes the model forcing factor affecting 
intensity of the expansion (limits to the $P$ variation for the REP 
simulations being studied here are determined in Section~\ref{sec.6.1}).
The conditions assumed result in development of the areas covered within
right half-plane from the IS (already at first stages of the REP, 
all the realizations evolve from the identical IS into actual finite random 
set, FRS defined in Section~\ref{sec.1}). The resulting REP tends to 
stable bi-state represented by the MMS patterns that do not grow further 
(see Sections~\ref{sec.5.1.1}, \ref{sec.5.3}). The schemes of local covering 
used to simulate the REP result in various forms of the covered area in the 
REP realizations, $A_i$. These can be sets of more or less dispersed rugged 
islands (see Fig.~\ref{fg.2}d, \ref{fg.3}) and this diversity is specific to 
modeling the {\itshape finite-size effects}.
\par
This specific feature of the REP realizations is revealed by comparison to a 
realization $B_i$, with $i \leq M$, of the spreading percolation (SPP) 
initiated from the same IS and expanding within right half-plane ~(see 
Fig.~\ref{fg.2}d). The corresponding assignment of the site covering 
probabilities for the SPP is
\begin {equation}
P_1=P_6=0 \; , \; P_7 \equiv 1, \; P:=P_2=P_3=P_4=P_5 \leq 1 .\label{eq.2}
\end {equation}
In course of the SPP evolution, a site
$\xi \in S_x$, with  $x \in B_i(T)$, is covered for the $B_i(T+1)$ if
$P_x(\xi) \geq r_x(\xi,\, i)$ with $\xi \in \{x,y_1,y_2,...,y_6\}$.
Thus, for the SPP, covering a site at a stage $T$ implies its covering at $T+1$
that is not the case for the REP (see Eqs.~(\ref{eq.1}) and~(\ref{eq.2})).
Let us note that SPP is used here as simplest process, without modeling the 
F-SE and being comparable to the REP, which enables us to emphasize effects 
attributed to the F-SE modeling in the REP.
\subsection {\label{sec.4.2}Diversity parameters $N$ and $M$} 
For reliability of the systematic study of the REP, one requires comparing 
results of simulations obtained for extensive series of the parameter values 
$P$, in a ~spatial ~domain ~$\chi \!=\! \{x_k, \; k=1,2,...,N\}$ with 
sufficiently large number $N$ of elements (nodes) and for appropriate number 
$M$ of the process realizations $A_i \; , \; i=1,2,...,M$. 
\par
The number $M$ of sets $A_i$ constituting the ensemble (FRS) parameterizes 
degree to which diversity of the FRS manifests itself in features observed in 
sequence of the mean expected patterns, $\int \! F(T)$ (i.e. in the 
sequence of MMS).
The value, $M_0$  limiting from below the number of realizations $M$
being moderate assures that all the realizations have no site covered in common
at any stage, $T>2$ of the random expansion process simulated.
Number $N\!\geq\!N_0\!\approx\!18000$ of sites constituting the space and
$M\!\geq\!M_0\!\approx\!100$ allow sufficient 
spatial-specific and ensemble-specific diversity modeled by respective
computing the pseudo-random matrices $\|r\|_{N \times 7 \times M}$. 
The $M$ different matrices having the size
$N \times 7$ and constituted by pseudo-random values are generated,
$r_{x_k}(\xi_j,i)$ with $k=1,2,...,N$, $\xi_j=y_j$ for 
$j=1,2,...,6$, $\xi_7=x_k$ and $i=1,2,...,M$.
Assessment of the upper bound $M_1$ to the $M$ being moderate results from this
investigation and is eventually presented in Section~\ref{sec.6.2.3}.
The spatial domain, $\chi$ should be sufficiently large to include each 
realization
at latest required stage $T_{mx}$ of the REP development,
\begin {equation}
{\mathop \bigcup_x} S_x \subset \chi \; \; \; \forall x \in 
base[F(T)]= \mathop {\bigcup_{i=1}^{M}} A_i(T)\; \; \; \forall T \leq T_{mx} .
\label{eq.3}
\end {equation}
The set, $base[F(T)]$ will be called {\itshape base} of the FRS
at the stage $T$ of the REP development; the symbols $\bigcup$ and $\subset$ 
denote union and inclusion of sets whereas $\forall$ is read "for all". 
Here, we have found that changes in number $N$ only,
$N_0\!\leq\!N\!\leq\!16\!\cdot\!N_0$
do not affect the simulation results remarkably.
\subsection {\label{sec.4.3}Lattice order projecting efficiency {\large 
$\alpha$}}
The finite discrete space $\chi$ is an example of a compact space that can be 
covered by finite number of open sets having  a structural property in common, 
which is then inherited by whole the space (e.g. \cite {16}. Here, this is the 
form of a regular hexagon. Extensive discussion of modelling 
the lattice order projecting efficiency has been presented by us previously 
\cite {17}. There, we have revealed the ways of achieving isotropy of the 
symmetrical spreeding on the hexagonal lattice, however, only one of them 
allows regulating {\itshape smoothly} the degree to which the lattice order is 
projected onto a process of the lattice covering. This way is followed here:
One can select randomly a fraction $\alpha$ of 
sites in $\chi$, for which covering all their six nearest neighbors is 
allowed. From the rest of sites, one selects randomly a half 
of them, $(1-\alpha)/2$ to allow covering
only odd sites within their closest vicinities $V_x$ (see Fig.~\ref{fg.2}a). 
For the remained half, one allows covering only even sites within the $V_x$.
The symmetry and mutual situation of the odd and even triplets within 
the $V_x$ result in mutual neutralization of their 
contributions to the inheritance effect. This can be observed for clusters 
constituted by sufficient number of sites covered. 
Eventually, the $\alpha$, $0\! \leq \! \alpha \! \leq \! 1$, parameterizes 
efficiency of projecting the order of regular hexagonal array onto patterns 
resulting from the REP development. 
\section {\label{sec.5}Characteristics of the REP development}   
\subsection {\label{sec.5.1}Characteristics of the process advancement}
\subsubsection {\label{sec.5.1.1}Reference to finiteness and discreteness}
After few initial stages, advancement in the REP development (either ITP or 
DTP) is characterized by monotonic growth in the average number of nodes 
covered in a realization,
\begin {equation}
I_{av}(T)= {1 \over M}\mathop {\sum_{i=1}^M} I_i(T) ,\label{eq.4}
\end {equation}
~with $I_i(T)\!=\! \mu \{A_i(T)\}$,  and in the number of nodes 
constituting the base, 
\begin {equation}
I_{bas}(T)=\mu\{bas[F(T)]\}\label{eq.5}
\end {equation}
~as well as by stepwise growth in the number of nodes of the MMS 
representation,
\begin {equation}
I_{mms}(T)=\mu\{\int \! F(T)\} .\label{eq.6} 
\end {equation}
~Here, $\mu \{.\}$ denotes the number of elements of the finite set $\{.\}$
~(see previous Sections for other denotations). 
The REP finishes at a stage 
$T_s$, until which all the $M$ realizations have finished their development
~(see Fig.~\ref {fg.3}a). 
A stage $T_a$, since which $I_{mms}(T)$ does not grow further, identifies
beginning of the REP {\itshape advancement} period.
Realizations $A_i(T)\; , \;i=1,2,...,M$ 
can finish their development at various stages $T \leq T_s$.
This variety of the realization evolutions results in $I_{bas}(T)$ growing much
stronger than $I_{av}(T)$ for $T\!>\!T_a$ (see Fig.~\ref{fg.3}).
In virtue of very small reductions in realizations in all REP simulations 
reported here, approaching the $T_s$ can be indicated by the $I_{bas}(T)$ and
$I_{av}(T)$ becoming constant.
\par
One may observe specific distribution, $I_{bas}(T)$ in the beginning of the 
period following the $T_a$ which corresponds to forms of not overlapping parts 
of few growing realizations. For example, a bend in the $I_{bas}(T)$, which 
can be observed around a stage $T_{bend}$ occurring close to the $T_a$, 
corresponds to change in form of increments in the base  from
rather uniform front like advancement to few narrow separated tongues
preserving direction of the expansion ~(see Fig.~\ref{fg.3}b). 
\par
We have found that smooth transition of the $I_{av}(T)$, from strong growth to 
very weak increments, indicates finishing of the evolution for major part of 
the realizations. These realizations, $A_i(T)$ finish their development rather 
abruptly within a period identified by the series of few stages which can be 
indicated conventionally by a central stage, $T_{bend}(i)$ of this period. 
This is indicated by a bend in the distribution $I_i(T)$. 
The $I_i(T)$ grows strongly until the bend and becomes constant for the 
following stages. Diversification of the values $T_{bend}(i)$ within the set
corresponding to the $M$ realizations and lack of the bend in $I_i(T)$ for a 
number of them result in the smooth distribution of the $I_{av}(T)$.  
\subsubsection 
{\label{sec.5.1.2}Specific effect of discreteness - steps in $I_{mms}(T)$} 
The growing length of steps in the $I_{mms}(T)$ evidences approaching the 
$T_a$ ~(see Fig.~\ref{fg.3}). We explain appearance 
of the steps in the $I_{mms}(T)$ with reference
to the method of determining the MMS (see Appendix~\ref{ap}).
\par 
At a stage $T$, one determines the $I_{av}(T)$ and pairs of sets,
$\Phi_i(T)$ or ~$\phi_i(T)$ constituted by such covered sites that a number of 
~$k \geq i$ ~or ~$k > i$ ~realizations have them in common, respectively. For 
the numerical criterion  $h=i/M$ used to determine the MMS, one selects such 
value $i$ that with this $i$ the inequalities
$\mu \{\phi_i(IT)\} \leq I_{av}(T) \leq \mu \{\Phi_i(T)\}$ are obeyed. 
A number of such covered sites that more than $i=const$ realizations have them 
in common can grow with $T$ for certain period. Therefore,
the values $\mu \{\Phi_i(T)\}$ and $\mu \{\phi_i(T)\}$ can also grow with
the $i=const$ to satisfy the noted inequalities when $I_{av}(T)$ increases 
with $T$. At certain stage of the following evolution, increment
in contribution to $I_{av}$ from less overlapping branches of the realizations
requires selecting smaller value $i$ to obey these inequalities. This change 
in the $i$ causes jump of increment in the $I_{mms}$ as stronger as larger is 
the contribution from not overlapping branches of the realizations. The period 
with $i=const$ is as longer as slower is growth in $I_{av}$ and as slower is 
growth in contribution to the increment in $I_{av}$ from dispersed in space, 
less overlapping branches of the realizations. This explains growth in length 
of the subsequent periods corresponding to subsequent values $i=const$, 
respectively. In virtue of the effects indicated by the steps in $I_{mms}(T)$, 
we will call the corresponding time intervals the {\itshape accumulation 
periods}.
\par
Thus appearance of the steps results from discreteness of the realization 
ensemble and their significance can be attributed to the remarkable share 
which not overlapping extended branches of the realizations have to the 
$I_{av}$ in the REP with the moderate $M$.
\par 
At earlier stages of the evolution, influence of the various factors onto the
$I_{mms}(T)$ is more complex and indicates significant contribution from
overlapping branches of the realizations.
\subsection {\label{sec.5.2}Structural characteristics}   
Characteristics of {\itshape local} clustering which have been collected 
within whole the space $\chi$ provide information about the pattern 
accomplished within the $\chi$ at a stage $T$. For that pattern, one may 
define a probability  ${\mathcal {P}}_n$ of finding  a cluster constituted by 
exactly $n$ sites covered within closest neighborhood $S_x$ of a site $x \in 
\chi$. We express it as ${\mathcal {P}}_n={\mathcal {R}}_n/{\mathcal {N}}_b$
with ${\mathcal {N}}_b$ being number of sites within the smallest rectangle
circumscribing the $base[F]$ and ${\mathcal {R}}_n$ being number of these
sites $x$ with the $S_x$ containing not less than one cluster constituted by 
exactly $n$ sites covered  $(2 \leq n \leq 7)$; note that using the form of a 
rectangle is just a convenient choice. It is easy to see that
occurrence of the $n$-site cluster within $S_x$ does not allow occurrence 
of a separate cluster constituted by different number, $n_1 \neq n$ and
$n_1 \geq 2$, of sites within this $S_x$. 
\par
Let us imagine a free gliding window $S$ congruent with $S_x$ which appears at 
a site $x$ if the $S_x$ coincides with this window ~(see Fig.~\ref{fg.2}a). We 
follow an approach known from the information theory \cite {18}  and consider 
the window $S$ as an information channel. A pattern that can be  observed 
within the $S$ can be considered as the information carrier passing through 
this channel. Distinct clusters composed of certain number $n$ of sites are 
independently transfered by the carrier through this channel while the free 
window $S$ appears once at each site within the smallest rectangle 
circumscribing the $base[F]$. Accordingly, one may consider the value,  
\begin {equation}
H_n = - {\mathcal {P}}_n \cdot ln {\mathcal {P}}_n \label{eq.7}
\end {equation}
as contribution from passing the $n$-site cluster to mean expected 
amount of information that can be transferred by the carrier through the 
channel $S$. The sum, $H = - {\mathop {\sum_{n=2}^7}} H_n$ is used here as 
reference scale for the $H_n$. Distributions of these relationships against 
the $T$, not just their values at $T$ fixed, characterize the evolution of the 
MMS or $A_i$ realization patterns. They are called here, for brevity, the  
{\itshape entropic characteristics}. We denote these values determined for the 
MMS pattern as, $H_{mms}$, $(H_n)_{mms}$, $(H_n/H)_{mms}$.
The values, $H_i$, $(H_n)_i$, $(H_n/H)_i$ are determined for a distinct 
realization $A_i$ whereas $H_{av}$, $(H_n)_{av}$, $(H_n/H)_{av}$ 
denote mathematical expectations of the entropic characteristics determined 
for every realization of the FRS.
\begin{figure*}
\includegraphics{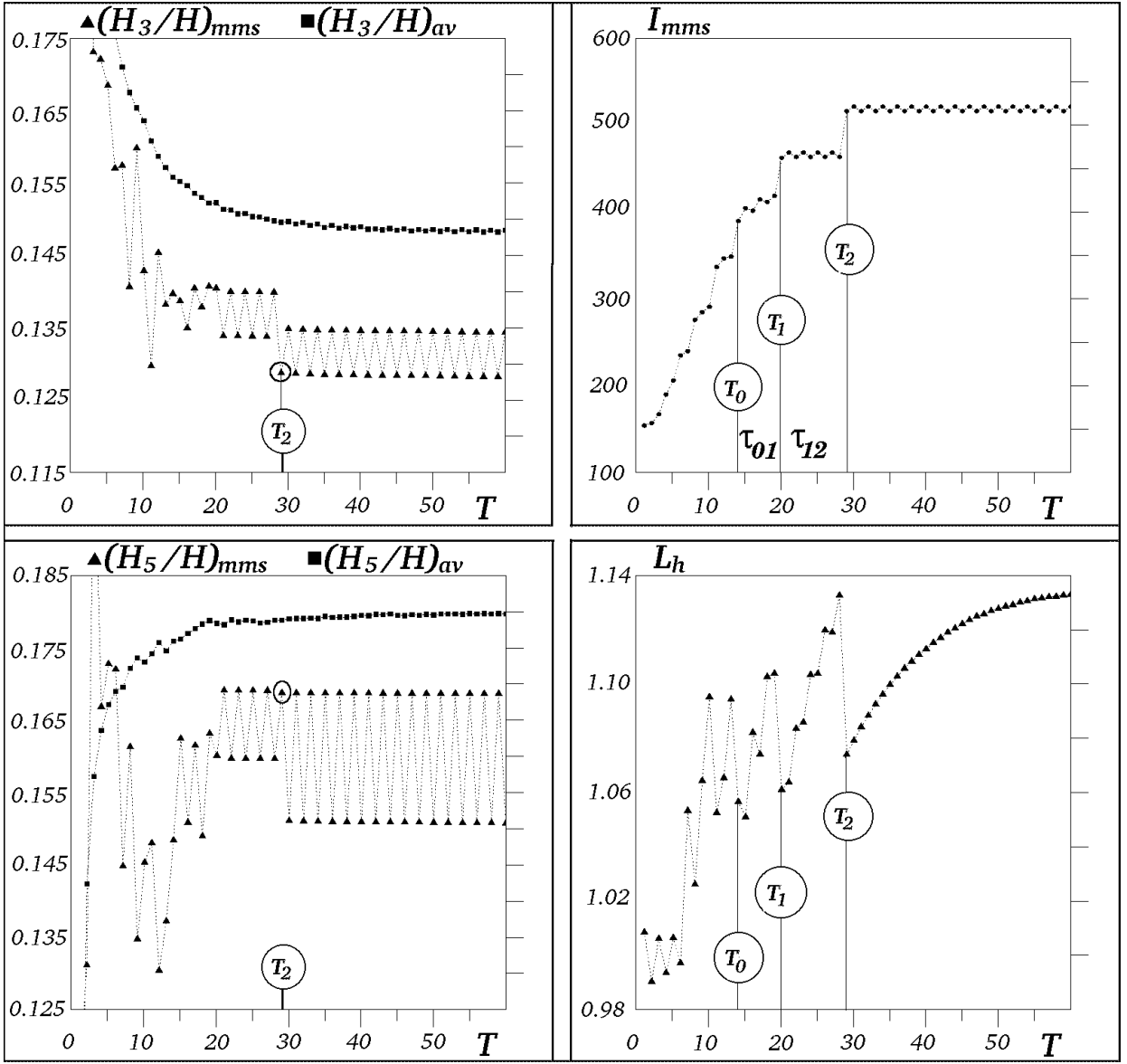}%
\caption{\label{fg.4} 
Characteristics of structural bi-state stabilization in the MMS development
for example of the REP-ITP with 
{\large $\alpha$} $={1 \over 3}$, $M=M_0\!=\!100$, $P=0.1311$.
The same value $T_2$ is identified in all graphs $(H_n/H)_{mms}(T)$ with
$2\!\leq\! n\!\leq \! 7$ ~(here, only the examples are shown).
}
\end{figure*}
\subsection {\label{sec.5.3}Features of MMS evolution}
\subsubsection {\label{sec.5.3.1}Structural bi-state}
Accomplishment of the structural bi-state in course of a number of initial
stages of the REP development is specific feature of modeling the finite-size
effects. Appearance of the bi-state in the MMS development is well illustrated 
by the graphs of the ~~$(H_n/H)_{mms}$ ~~(see Fig.~\ref{fg.4}). 
Saw-shaped distribution of the $(H_n/H)_{mms}$ against $T$ depicts these 
values at subsequent stages $(T, \; T+1)$ as corresponding to states of the  
bi-state. Situation of the points representing the subsequent pairs, 
$\{(H_n/H)_{mms}(T) \;,\; (H_n/H)_{mms}(T+1)\}$ on parallel lines  for an 
accumulation period detects structural bi-state stability of the MMS 
development within this period. Then, a pair of neighbor stages computed, 
$(T, \; T+1)$ corresponds to a single physical stage when the system may be in 
one as well as in the other structural state.
Result of this type has been reported  by us previously \cite {9} as the one 
attributed to modeling the finite-size effects and, then, its relevance to 
available experimental observations has been noted.  
\par
The REP simulated accordingly to the REP-ITP and REP-DTP reveals features of 
the bi-state stabilization. Let us recall that patterns of covered sites, which
are being accomplished within whole the spatial domain in course 
of the REP-ITP, may be considered as resulting from processes of covering the 
sites accordingly to both the schemes, REP-DTP and REP-ITP-P (see 
Section~\ref{sec.3.3}). However, the REP simulations performed accordingly to 
the REP-ITP-P scheme only, resulted in appearance of the 
bi-state with no features of its stabilization. This shows that sufficient 
contribution from the REP-DTP to the REP-ITP is the necessary condition of the 
bi-state stabilization for the REP-ITP. This condition is satisfied for all 
the REP-ITP simulations  reported here (effect of the contribution which 
REP-DTP has to REP-ITP is elucidated in Section~\ref{sec.6.2.2}). 
\begin{figure*}
\includegraphics{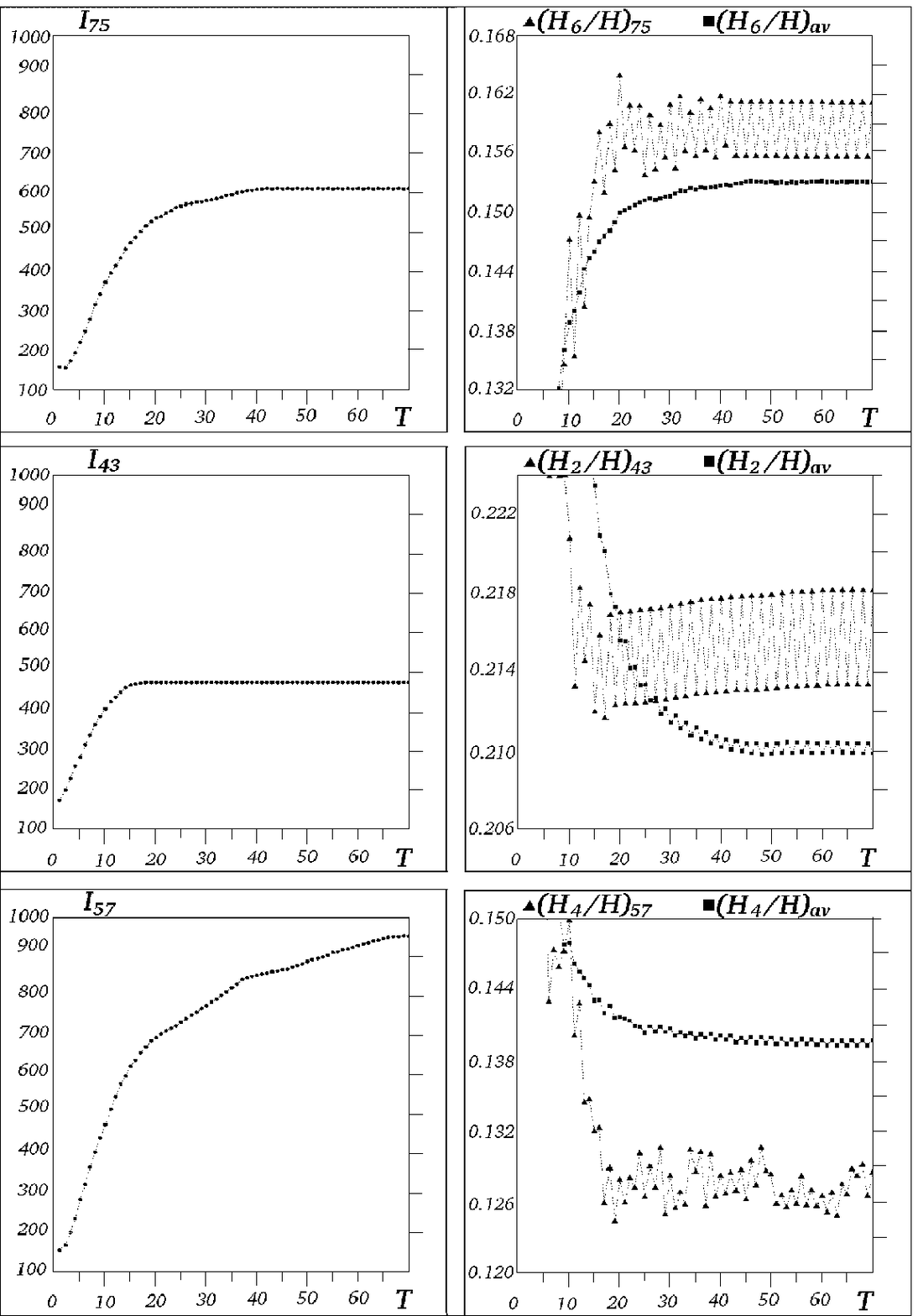}%
\caption{\label{fg.5} 
Examples of characteristics representing three recognized classes of 
originating the structural bi-state and its stabilization in a realization 
development. For
this example of the REP-ITP with {\large $\alpha$} $={1 \over 3}$, 
$M=M_0\!=\!100$, $P=0.1311$ (then, $T_0=14$, $T_1=20$, $T_2=29$), of about 
$35\%$ of the realizations $A_i(T)$ have appeared in stable structural
bi-state before the stage $T=T_2$.
Observe that appearance of the bi-state close to $T_0=14$,
at $T=15>T_0$ ~$(i\!=\!75)$ or at $T\!=\!13<T_0$ ~$(i\!=\!43)$ corresponds to 
its
stabilization and becoming the $I_i(T)=const$ at stages $T\!=\!42>T_2$ 
~$(i\!=\!75)$
or $T\!=\!18<T_1$ ~$(i\!=\!43)$. Here, later origination of the bi-state in
development of the realization $i\!=\!57$ corresponds to absence of the 
stabilization
in the $A_{57}(T)$ before $T\!=\!70 \gg T_2$.
}
\end{figure*}
\begin{figure*}
\includegraphics{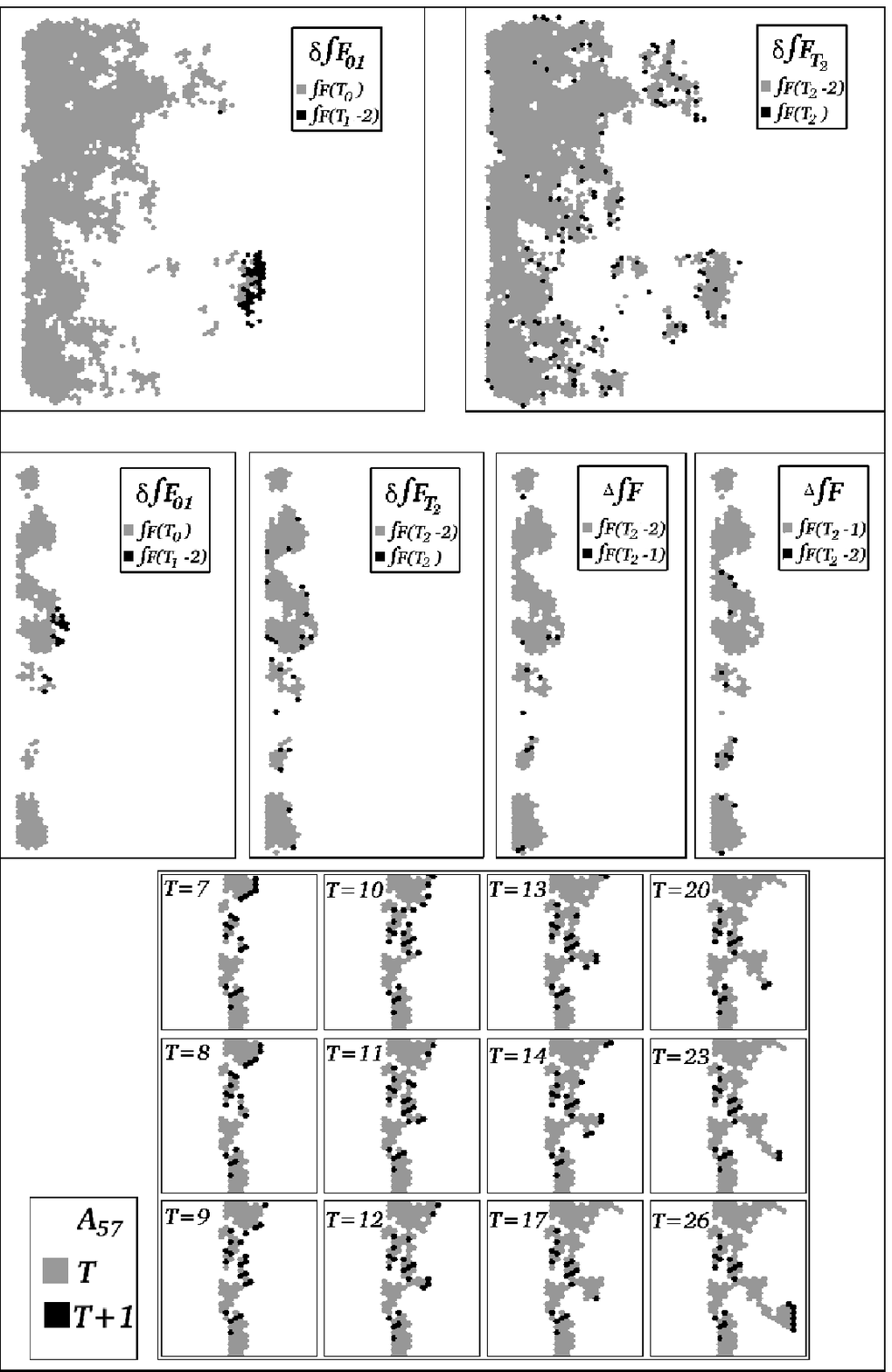}%
\caption{\label{fg.6} 
Pairs of MMS patterns imposed one on the other to show characteristic 
increments in
the MMS in course of the REP evolution. For example of a region in one REP 
realization $A_{57}$, the pairs of imposed
patterns are shown to reveal appearance of the bi-state. 
In all figures, 
bright pattern is imposed on the dark one. 
All figures have been obtained by simulating the REP-ITP
with {\large $\alpha$} $={1 \over 3}$, $M=M_0\!=\!100$.
The top pair of patterns results from the simulation
with high $P=0.18$
~(then, $T_0\!=\!82 \;,\; T_1\!=\!92 \;,\; T_2\!=\!101)$
whereas all other figures with low $P\!=\!0.1311$ 
~(then, $T_0\!=\!14 \;,\; T_1\!=\!20 \;,\; T_2\!=\!29)$.
}
\end{figure*}
\subsubsection {\label{sec.5.3.2}Bi-state stabilization events, $T_0$, $T_1$, 
$T_2$}
Appearance of the bi-state stability at the end $T_e$ of certain accumulation
period is identified if this effect encompasses the computed sequence from 
$(T_e \!+\!1\!-\!2\cdot t)$ to the $T_e$, with $t\!\geq\!2$,  for the
$(H_n/H)_{mms}(T)$ for all $n$, $2\! \leq \! n \! \leq \! 7$. The bi-state  
stability accomplished is preserved for all the following accumulation 
periods. One may observe such accumulation period that its end indicated by 
the $T_e$ is the first computed stage in course of the evolution, for which 
the conditions just mentioned are obeyed. Then, the stage $(T_e\!+\!1)$
is denoted as $T_2$ and corresponds to the first jump from one 
structurally stable bi-state to other such bi-state (see Fig.~\ref{fg.4}). 
This corresponds to a jump in the $I_{mms}(T)$ between corresponding
neighbor steps, however, the $I_{mms}(T)$ cannot replace the 
$(H_n/H)_{mms}(T)$ in unambiguous detecting the $T_2$. The accumulation period
preceding the $T_2$ begins at the stage denoted as $T_1$ whereas the one 
preceding this begins at the stage denoted as $T_0$. Having given the $T_2$, 
both the $T_1$ and $T_0$ can be easily detected as stages corresponding to 
local minima in distribution of a helping characteristic 
$L_h(T)=1+(H_{av}-H_{mms})/H_{av}$ ~(see Fig.~\ref{fg.4}) or, for higher 
values of $P$, as instants of jumps between steps in the $I_{mms}(T)$.
\par
The distributions, $(H_n/H)_{i}(T)$ with $1 \leq i \leq M$, reveal that $T_0$ 
or $T_1$ is within a short sequence of the stages since which the realizations 
can start accomplishing the bi-state not stable yet or can accomplish stable 
structural bi-state, respectively (see Fig.~\ref{fg.5}); hereafter we will use 
also the phrase {\itshape becoming bi-stable} as abbreviation to 
{\itshape accomplishing the stable bi-state}. Significant part of 
realizations becomes bi-stable until the $T_2$, however, a number of them  
becomes bi-stable very late after the $T_2$ and they begin accomplishing the 
bi-state close to the $T_1$.
\par
Eventually, one may consider the triplet, $T_2$, $T_1$, $T_0$ as values
characterizing evolution of the MMS pattern in the REP. Their distributions 
against $P$ can reveal possible regular changes in these evolutions with 
varying $P$.
\subsubsection {\label{sec.5.3.3}Characteristic increments in the MMS 
development}
Sets being increments in the $\int\!F(T)$ for the
accumulation period or at the jump between these periods are to be determined 
for the same state of the bi-state (see Fig.~\ref{fg.6}). 
These sets are denoted as $\delta\!\int\!F$ and identified by the pair of 
indices $l,k$ or $lk$ for the respective accumulation period whereas index 
$T_j$ is used to indicate sets being increments at a jump at the $T_j$.
Accordingly, the pairs $01$ or $12$ indicate sets being increments in the 
$\int\!F(T)$ for the accumulation periods, from $T_0$ to $T_1$ or from $T_1$ 
to $T_2$, respectively. The accumulation periods, $\tau_{l,k}$ and instances 
of the jump events $T_j$ are calculated for each state of the bi-state.  
We calculate length of the
accumulation period between simulated stages for 
$l \in \{0,1\}$ and  $k=l+1$ as,
$\tau_{l,k} = t_{k}-T_l+1$, with $t_{k} = T_k-1$ or
$t_{k} = T_k-2$, so that $t_{k}$ and $T_l$ are both odd
or they both are even. One considers increments 
$\delta \! \int \! F_{l,k}$ 
corresponding to the $\tau_{l,k}$: between $T_l$ and
$t_{k}$ for one state and between $T_l+1$ and $t_{k,A}$ for the other, with
~$t_{k,A}=t_k-1$ ~if ~$t_k=T_k-1$ 
~or ~$t_{k,A}=t_k+1$ ~if ~$t_k=T_k-2$. 
The increments $\delta \! \int \! F_{l,k}$ are determined accordingly to
Eq.~(\ref{eq.8}). The nonnull reductions, $\delta \! \int \! F_{k,l}$
determined in accordance with Eq.~(\ref{eq.9}) may accompany these increments.
Changes in forms of these increments and reductions with varying the $M$, 
{\large $\alpha$} and $P$ are considered in Section~\ref{sec.6.2.3}. 
Increments in $\int \! F$ corresponding to a state of the bi-state at
the jump in MMS growth at $T_j$ are sets of sites scattered throughout whole 
the $\int \! F$ and they are not accompanied by a reduction in the MMS. They 
are denoted as $\delta \! \int \! F_{T_j}$ and determined accordingly to 
Eq.~(\ref{eq.10}). Differences $\Delta \! \int \! F(T,T+1)$ 
between MMS corresponding to the two states of the bi-state
are given by Eq.~(\ref{eq.11}) and they
have also the form of a set of covered sites dispersed throughout the 
$\int \! F$. All the sets just noted are given by the following relationships,
\begin{widetext}
\begin {equation}
\delta \! \int \! F_{l,k} \in \left \{\int \! F(t_k) \! \setminus \!
\int \! F(T_l) \; , 
\; \int \! F(t_{k,A}) \! \setminus \! \int \! F(T_l+1) \right \},
\label{eq.8}
\end {equation}
\begin {equation} 
\delta \! \int \! F_{k,l} \in \left \{\int \! F(T_l) \! \setminus \!
\int \! F(t_k) \; , \; \int \! F(T_l+1) \! \setminus \! \int \! F(t_{k,A}) 
\right \},
\label{eq.9}
\end {equation}
\begin {equation} 
\delta \! \int \! F_{T_j} \in \left \{\int \! F(T_j) \! \setminus \!
\int \! F(T_j-2) \; , \; \int \! F(T_j+1) \! \setminus \! \int \! F(T_j-1)
\right \},\label{eq.10}
\end {equation}
\begin {equation} 
\Delta \! \int \! F(T,T+1) = \left \{\int \! F(T) \! \setminus \! 
\int \! F(T+1) \; , \; \int \! F(T+1) \! \setminus \! \int \! F(T) \right \},
\label{eq.11}
\end {equation}
\end{widetext}
(here, $A \setminus B$ denotes elements of a set $A$ which 
are not elements of the set $B$).
The form of increment in the MMS
depicted in Fig.~\ref{fg.6} for example of the 
$\delta \! \int \! F_{T_2}$ as well as  $\Delta \! \int \! F(T,T+1)$
will be identified as {\itshape scattered increment}.
\par
Sequences of patterns representing evolution of distinct realizations $A_i(T)$
reveal nature of accomplishing the bi-state and thus elucidate emerging the
characteristic forms of the increments in MMS in course of the REP 
development. Example of such sequence is presented in the array of patterns 
shown for $A_{57}$ in Fig.~\ref{fg.6}. Specifically developed form of the 
realization within its central region, distanced from the fronts, results from 
accomplishment of the bi-state. No bias from this specific feature in 
evolution of a realization has been observed either for REP-DTP or REP-ITP 
simulated here.
\begin{figure*}
\includegraphics{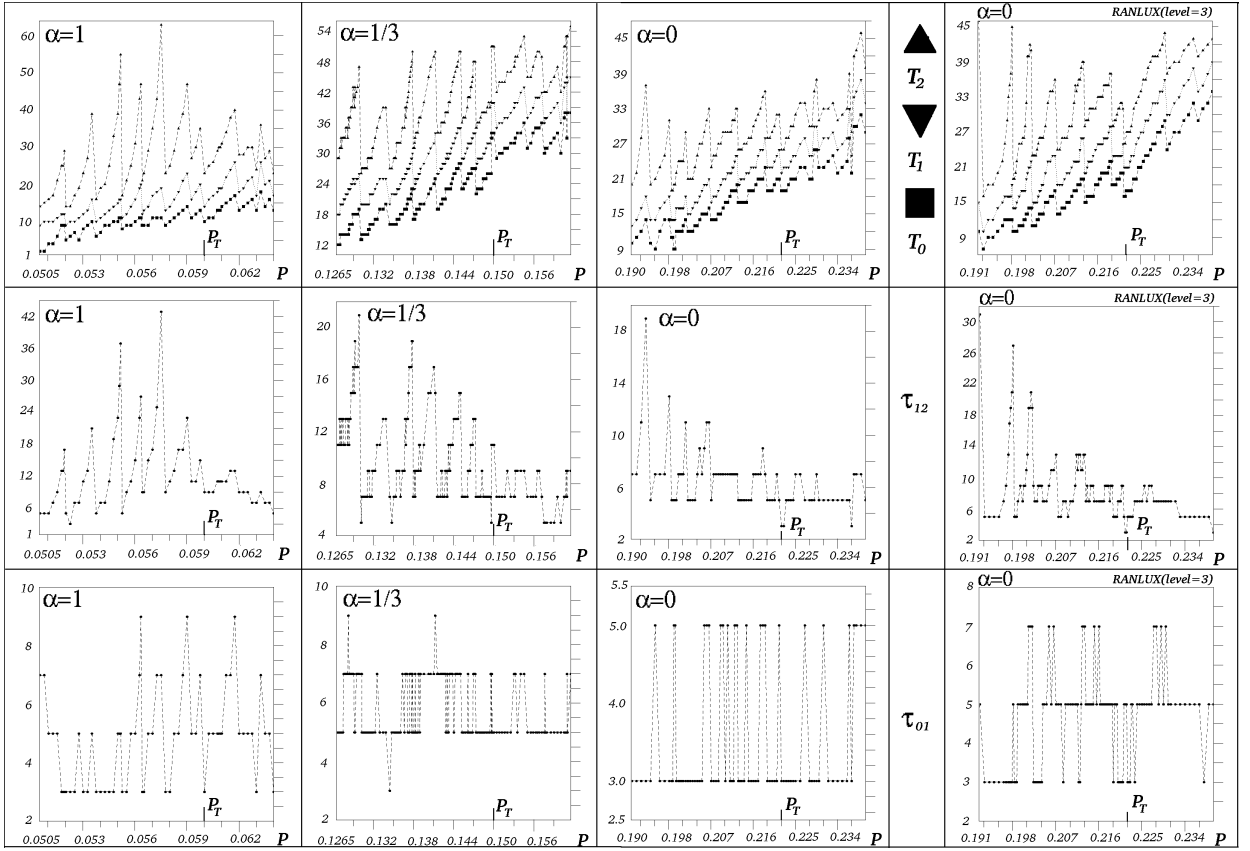}%
\caption{\label{fg.7} 
Regular variations of the MMS bi-state stabilization event characteristics 
with $P$ and {\large $\alpha$}. All the REP-ITP simulations have been
done with $M=M_0$, the same pseudo random matrix, 
$\|r\|_{N_0 \times 7 \times M_0}$ and space of $N=N_0$ sites.
The same pseudo random number generator RANMAR has been employed to compute 
this matrix and to satisfy the condition, {\large $\alpha$} $<1$. Only the 
control distributions shown in the column most to right have been obtained by 
using the more advanced generator RANLUX (with luxury level=3).
}
\end{figure*}
\section {\label{sec.6}Results and discussion}
One searches for regularities in changes of the MMS evolution patterns $\int\! 
F(T)$ with varying the process parameters $P$, {\large $\alpha$}, $M$. 
Results of the previous section suggest that distributions, 
$T_0(P),\;T_1(P),\;T_2(P)$, $\tau_{01}(P),\;\tau_{12}(P)$ with $P$ varying
within an interval when {\large $\alpha$}, $M$ are kept constant may reveal
such regularities. The corresponding increments  $\delta \! \int \! F_{T_j}$,
$\delta \! \int \! F_{01}$,  $\delta \! \int \! F_{12}$ associate the
changes in the characteristics with features of the MMS evolution patterns.
Accordingly, one investigates these characteristic distributions with fixed 
parameters {\large $\alpha$} $\in \{0,{1 \over 3},1\}$ and $M=M_0\!=\!100$.
Dependence of the $\int\! F(T)$ on the number of realizations $M$ is 
investigated with the purpose to estimate the upper limit $M_1$ to the values 
$M$ being moderate.
\subsection {\label{sec.6.1}
Distributions $T_j(P)$, $\tau_{01}(P)$, $\tau_{12}(P)$} 
The three sets of distributions $T_j(P)$, $\tau_{12}(P)$, 
$\tau_{01}(P)$ have been obtained from the series of REP-ITP
simulations with  {\large $\alpha$} $\in \{1,{1 \over 3},0\}$,
$M=M_0$, in the same space $\chi$ 
($N_0 \approx 18000$ nodes) and by employing the same matrix 
$\|r\|_{N_0 \times 7 \times M_0}$ of pseudo-random numbers. 
The generator RANMAR \cite {13} has been employed to compute the  matrix 
$\|r\|_{N_0 \times 7 \times M_0}$ and to satisfy condition of the lattice 
order projecting efficiency 
{\large $\alpha$} $<1$ (see Section~\ref{sec.4.3}). Three triplets of 
graphs depicting the distributions $T_j(P)$, $\tau_{12}(P)$,
$\tau_{01}(P)$ with  {\large $\alpha$} $\in \{1,{1 \over 3},0\}$
are presented in Fig.~\ref{fg.7}. With the purpose to recognize an effect 
of using a particular pseudo-random number generator, they are compared to  
results of the REP-ITP simulation series repeated with employing the more 
advanced generator RANLUX(luxury level=3) \cite{14,15} 
instead of using RANMAR to compute the $\|r\|_{N_0 \times 7 \times M_0}$ 
and to satisfy the requirement  {\large $\alpha$} $=0$. 
This comparison reveals only quantitative differences that appear to have no 
effect onto inferences about qualitative features of certain regularities 
observed in the distributions. Therefore, the simpler generator RANMAR has 
been considered as sufficient to obtain reliable results for the purpose of 
this research. 
\par
\begin{figure*}
\includegraphics{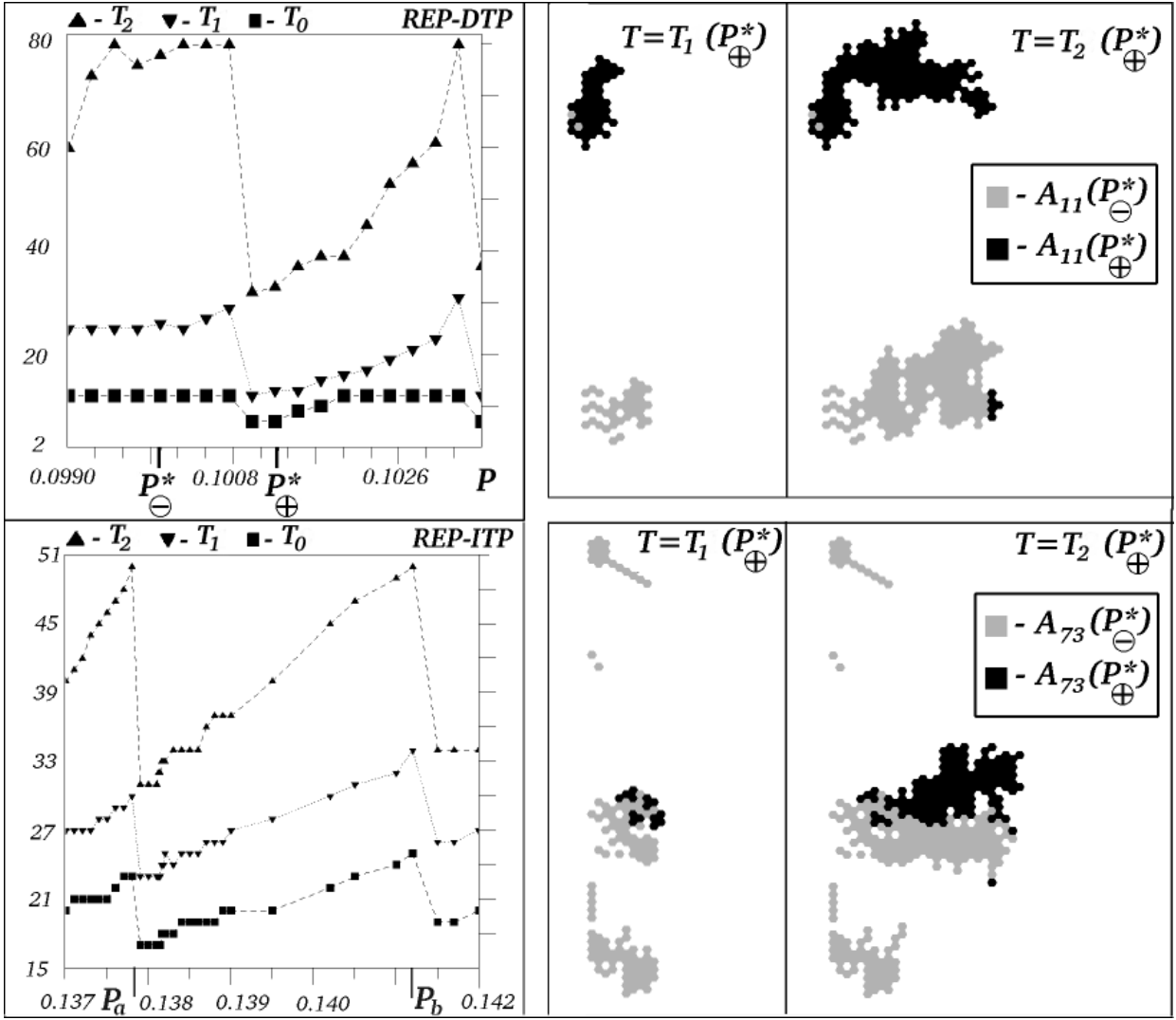}%
\caption{\label{fg.8} 
Spatial uniformization of contributions to the MMS from realizations of the REP
simulated with series of the $P$. This uniformization, which corresponds to 
appearance of the critical value $P^*$, is shown for example of the REP-DTP 
series with $M\!= \! 104\! \approx \!M_0$, {\large $\alpha$}$=1$ and
growing $P$ ~(here,
$P_{\ominus}^* \! = \! 0.10000\! < \! P^* \! < P_{\oplus}^* \! = \! 0.10125$).
The figures depict bright pattern imposed on the black one.
Regular character of distributions $T_j(P)$ within interval of $P$ between 
subsequent critical values, $P_a \;,\; P_b$ is specific to the REP-ITP; this is
shown for example obtained with
{\large $\alpha$} $={1 \over 3}$, $M=M_0\!=\!100$.
}
\end{figure*}

The interval of the values $P$ used to perform series of the REP-ITP
simulations has been selected specifically for each value of the 
{\large $\alpha$} $\in \{1,{1 \over 3},0\}$. The lower limit of the $P$
variations has been selected so as to obtain the REP-ITP evolutions revealing
clearly occurrence of all the three instants $T_0$, $T_1$, $T_2$. 
The upper limit has been assumed with the help of the $\tau_{12}(P)$. One may 
observe qualitative change in variation of the $\tau_{12}$ with growing $P$
starting from certain threshold value $P=P_T(${\large $\alpha$}$)$.
For $P>P_T$, the value $\tau_{12}(P)$ appears to vary more regular 
and within fixed interval (see Fig.~\ref{fg.7}). This enables us to estimate,
$P_T(${\large $\alpha$}$=0) \approx 0.222$ (for RANMAR and RANLUX),
$P_T(${\large $\alpha$}$={1 \over 3}) \approx 0.150$,
$P_T(${\large $\alpha$}$=1) \approx 0.060$, that correspond to the beginning
of certain interval $[P_a,P_b]$ within which the $T_j(P)$ is growing 
(see Fig.~\ref{fg.7} and Fig.~\ref{fg.8}; appearance of such intervals is 
elucidated in next Section~\ref{sec.6.2.1}). 
The $P_T(${\large $\alpha$}$)$ appears to be indicated also 
by specific distributions $T_j(P)$ within two neighbor subintervals, 
$[P_a,P_b]$  situated just to left of the $P_T$ and within the subinterval
just to right of the $P_T$. Eventually, the upper limit of the $P$ has been 
assumed so as to encompass not less than two subintervals $[P_a,P_b]$ 
following the $P_T(${\large $\alpha$}$)$. A number of test simulations 
performed accordingly to the REP-ITP scheme with the $P$ being substantially 
higher than the limits selected seem to corroborate  reliability of using the 
$P_T(${\large $\alpha$}$)$ to assess scope of the $P$ abscissa for this
investigation.
\par
Form of the graph $T_j(P)$ resembling quasi-periodical distribution as well as
occurrence of the $P_T(${\large $\alpha$}$)$ are specific to the 
REP-ITP and have not been observed for REP-DTP simulation series.
The distribution, $T_j(P)$ is much less regular for the REP-DTP simulated with 
the $M\!\approx\!M_0$ (see Fig.~\ref{fg.8}) and much more complex for the 
REP-DTP series computed with high values of the $M$.
\subsection {\label{sec.6.2}Structural pattern of changes in REP series}
\subsubsection {\label{sec.6.2.1}Appearance of intervals $[P_a,P_b]$} 
Interpretation of the distributions $T_j(P)$ requires comparing the 
corresponding realizations, $A_i(T)$ with the same $i\!<\!M$, in the REP 
simulation series of the same type, performed with close values $P$ 
constituting the growing series and with other parameters being the same for 
these simulations.
\par
One can observe that there are intervals of the value $P$ within which
a small increment in $P$ may result in covering some sites being
incipient elements of substantial branches developing in some realizations. 
In virtue of the $M$ being moderate, those branches can postpone occurrence 
of the $T_2$ (see also Sections~\ref{sec.5.1.2},~\ref{sec.5.3.2}). 
Further growth in $P$ may result in covering more sites constituting the 
incipient elements of the branches in larger number of realizations. 
Evolutions of the corresponding realizations obtained with close values of $P$ 
and for the respective accumulation periods reveal growth of the spots 
covered, increment in number of those spots and, eventually, spatial 
uniformization of their arrangement (see Fig.~\ref{fg.8}).
Then, one may observe growth in number of covered spots distributed along 
an interval being parallel to the initial chain (see the example of $A_{11}$ 
in Fig.~\ref{fg.8}) or expansion of the localized tongues also in this 
direction (see the example of $A_{73}$ in Fig.~\ref{fg.8}). 
The spatial uniformization of the realizations appears to be sufficient for
stabilization of the bi-state at an earlier stage of the evolution if $P$ 
becomes slightly greater than certain value $P^*$ (the $P^*$ appears as 
right-hand end of the interval alluded to previously; the $P$ just mentioned 
as being slightly greater than the $P^*$ is denoted as $P^*_{\oplus}$).
This earlier stage appears closer to the $T_1$ detected in the evolution
performed with $P$ being slightly smaller than $P^*$ ~(this value of $P$ is
denoted by $P^{*}_{\ominus}$). Thus the inequality, 
$T_2(P^{*}_{\ominus}) > T_2(P^{*}_{\oplus})$ is obeyed. Nature of the changes
just revealed explains repetition of the scenario for the REP evolutions with 
the $P$ growing further and thus appearance of subsequent intervals 
$[P_a^*,P_b^*]$ ~(they are denoted also simpler as $[P_a,P_b]$). 
\begin{figure*}
\includegraphics{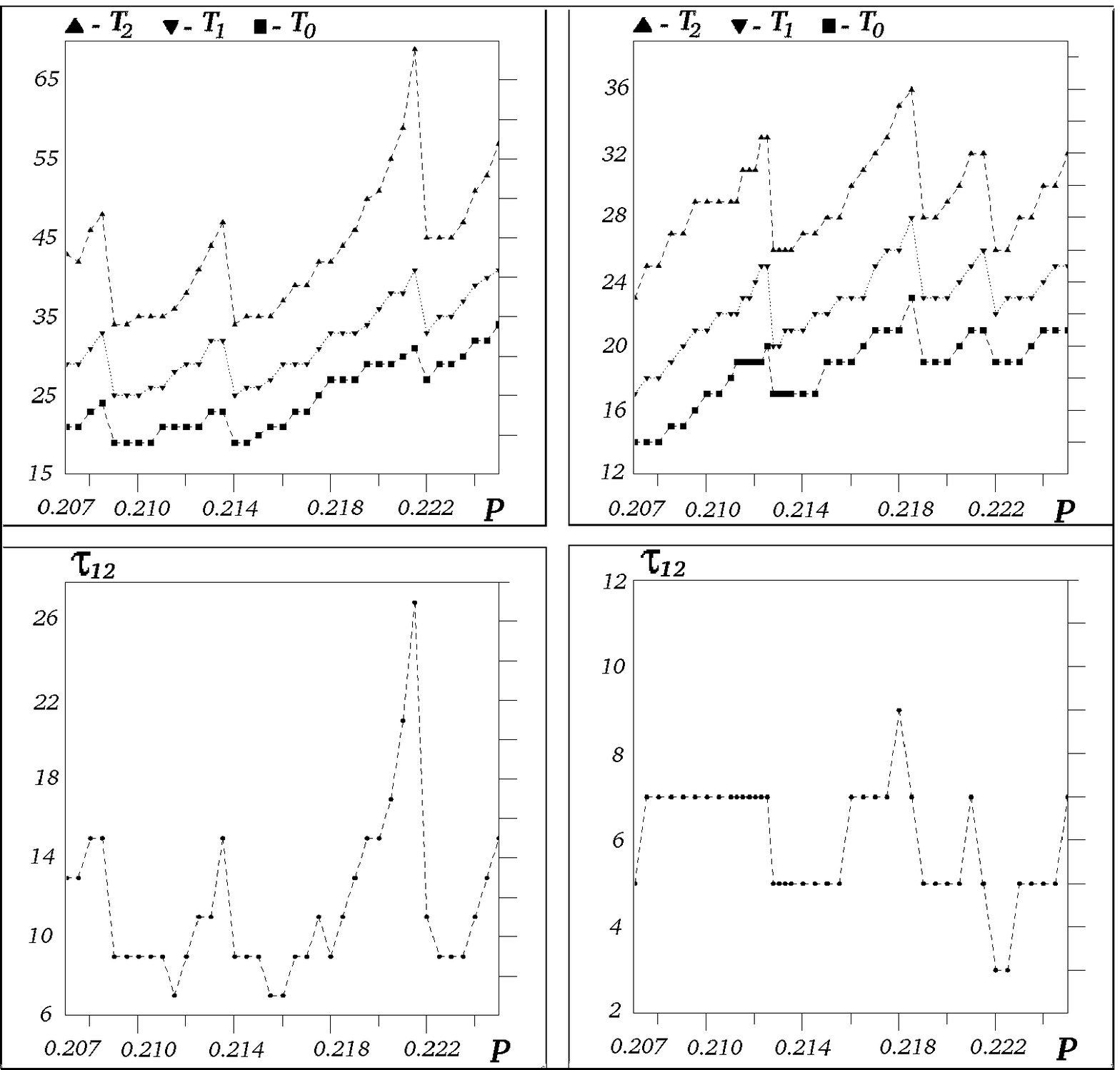}%
\caption{\label{fg.9} 
Different degrees of harmony between distributions $T_2(P)$ and $T_1(P)$ which
correspond to two ways of satisfying the condition of zero efficiency of the 
lattice order projecting onto REP (the example of the REP-ITP simulated with 
$M\!= \! M_0$). Figures in the left column depict results obtained with this 
condition satisfied by random assigning all the possible arrangements of sites 
whose covering is allowed within vicinities $V_x$ to all $x\in\chi$ . The 
right column depicts the results obtained
with  {\large $\alpha$}$=0$ satisfied with preserving symmetry in arrangements 
of sites whose covering is allowed in  $V_x$ (this method is used here; 
see Section~\ref{sec.4.3} and Ref.~\cite {17} for further explanations 
concerning the {\large $\alpha$}).
}
\end{figure*}
\subsubsection {\label{sec.6.2.2}Synchronisation effect of covering sites in 
REP-ITP}
The distributions $T_j(P)$ obtained from the series of the REP-ITP
simulations reveal variation resembling the quasi-periodical one and $T_2(P)$
grows rather monotonically within the intervals $[P_a,P_b]$
(see Fig.~\ref{fg.7} and Fig.~\ref{fg.8}). 
This is accompanied by systematic dependence of the
average length  $\mathcal {L}_P$ of intervals $[P_a,P_b]$ on the
{\large $\alpha$}: $\mathcal {L}_P(${\large $\alpha$}$)/\mathcal {L}_P(${\large
$\alpha$}$=1) \approx 3-2 \cdot${\large $\alpha$}.
Evolutions of realizations in the REP-ITP series simulated with the series
of $P$ growing within the $[P_a,P_b]$ toward the $P_b$ 
$(P\!\rightarrow\!P_b)$ result in covering larger number of sites dispersed in
the space. Then, more stages is required to accomplish sufficiently dense 
sets of the sites covered in the realizations to enable sufficient 
contribution from the REP-DTP to REP-ITP for stabilization of the bi-state 
(this contribution has been explained in Section~\ref{sec.5.3.1}). 
Further growth of the $P$, until $P_{b\oplus}$ is reached, results in denser 
covering the area in realizations $A_i(T)$ so that contribution from the 
REP-DTP to REP-ITP becomes more substantial and the stabilization occurs 
earlier. This may be interpreted as result of {\itshape synchronization} 
between site covering accordingly to the REP-ITP-P and REP-DTP schemes 
considered as components of the REP-ITP scheme.
\par
One may note following factors contributing to the regularity of the 
distributions $T_j(P)$. The REP-ITP differs from the REP-DTP in exploring the
larger number of site configurations within neighborhoods of sites covered at 
a stage $T$ while searching for pairs realizing transfers resulting in 
covering target sites at the stage $T+1$ (see Section~\ref{sec.3.3} and 
Fig.~\ref{fg.2}c). This results in comparable contribution of all the 
realizations $\{A_i\}_{i=1\div M}$ to the MMS in the REP-ITP simulations.
Preserving the symmetry in the arrangement of the odd or even sites 
whose covering is allowed within vicinities $V_x$ with 
{\large $\alpha$}$<1$ (see Section~\ref{sec.4.3} and Ref.~\cite {17}) 
facilitates more uniform covering the space in realizations 
$\{A_i\}_{i=1\div M}$ and, eventually, appears to improve harmony between 
the distributions $T_2(P) \;,\; T_1(P)$ within the intervals $[P_a,P_b]$ 
~(see Fig.~\ref{fg.9}).
\begin{figure*}
\includegraphics{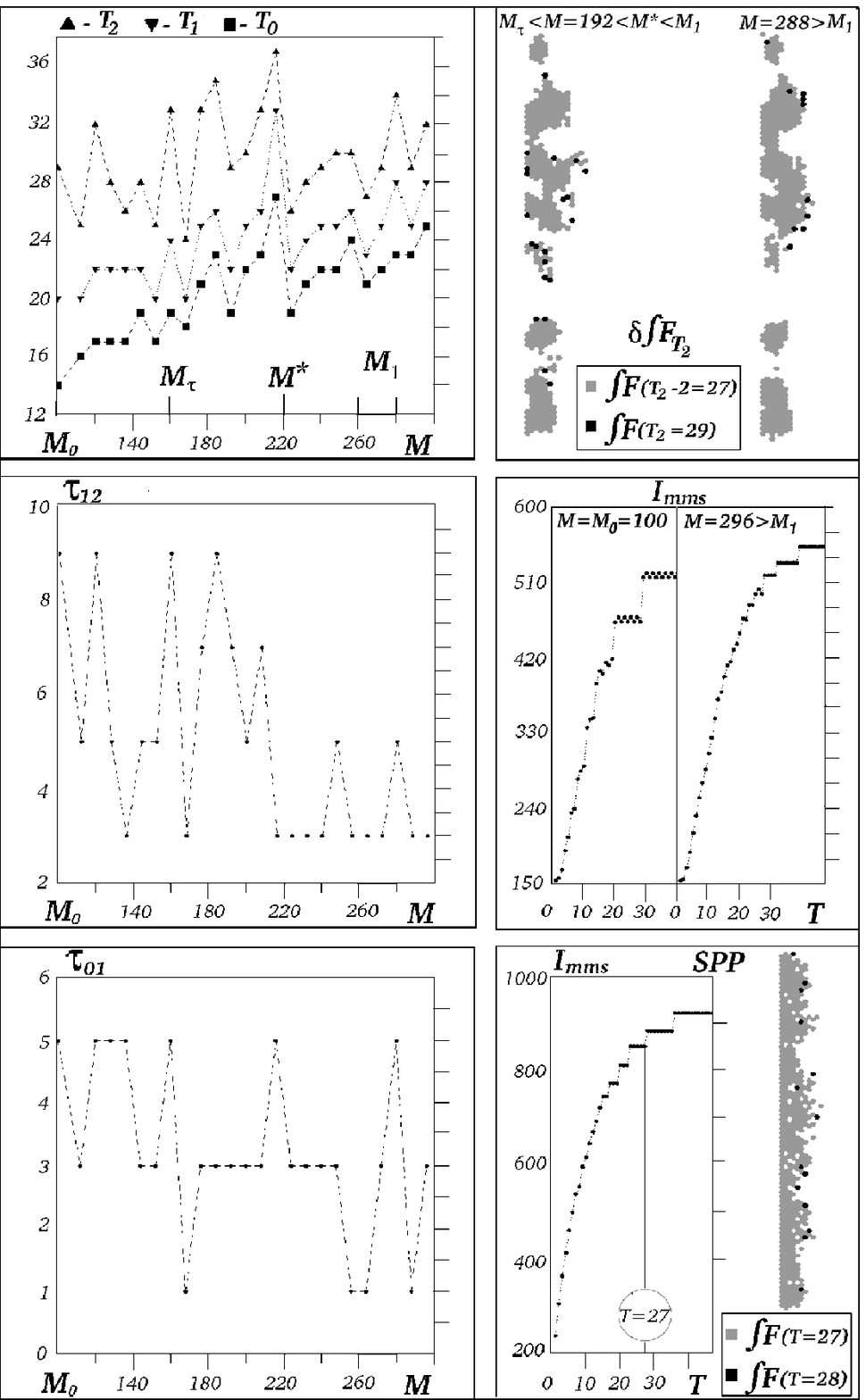}%
\caption{\label{fg.10} 
Identifying the upper bound $M_1$ of the $M$ being moderate for example of the 
REP-ITP series with $P\! =\! 0.1311$, {\large $\alpha$} $={1 \over 3}$. For 
this example, a pair of distributions, $I_{mms}(T)$ illustrates diminishing 
the effect of discreteness with growing $M$. 
Compare the increment $\delta \! \int \! F_{T_2}(M\! > \! M_1)$ to the
increment in MMS at jump in $I_{mms}$ appearing in course of the SPP evolution
~(example  of the SPP with $P\! =\! 0.475$, 
{\large $\alpha$} $={1 \over 3}$,  $M\! =\! M_0\!=\!100$ is depicted by
the indicated figure only). The figures depict bright pattern imposed on the
black one.
}
\end{figure*}
\subsubsection 
{\label{sec.6.2.3}Variations of MMS evolution patterns with $P$, 
{\large $\alpha$}, $M$}
Dependence of the increments, $\delta \! \int \! F_{01}$,
$\delta \! \int \! F_{12}$
on the simulation parameters $P$, {\large $\alpha$}
with $M=M_0$ is presented for the REP-ITP in Table~\ref{tab.1}. 
Note that there are no increments for the accumulation periods following the
~$\tau_{12}$ ~if ~$\delta \! \int \! F_{12}$ vanishes.
Occurrence of reductions, 
$\delta \! \int \! F_{10}$ and $\delta \! \int \! F_{21}$
accompanying increments, 
$\delta \! \int \! F_{01}$ and $\delta \! \int \! F_{12}$ in the MMS 
for the accumulation periods $\tau_{01}$, $\tau_{12}$ as well as 
characteristic differences between those increments (reductions) in the 
REP-ITP evolutions, with $P \rightarrow P_{a\oplus}$ and $P \rightarrow 
P_{b\ominus}$ for all $[P_a,P_b]$, are specific features of the REP-ITP
simulated with higher values of the {\large $\alpha$}; the reductions have not 
been observed for REP-ITP with small values of the {\large $\alpha$} and for 
the REP-DTP ~(see Tables~\ref{tab.1},~\ref{tab.2}).
\begingroup
\squeezetable
\begin{table*}[ht] 
\caption{\label{tab.1}
Dependence of increments in MMS on $P$ and {\large $\alpha$} for REP-ITP.
In the Tables, $Si$ denotes sedimentation-type increment (see Fig.~\ref{fg.6})
which can be solid or dispersed; symbol $\oslash$ identifies a null set or
the increment being a single site covered whereas $\sim\!\oslash$ shows that
the increment is constituted by not more than $2 \div 4$ not clustered covered 
sites (see Sections~\ref{sec.5},~\ref{sec.6.1},~\ref{sec.6.2} for 
other denotations). The results presented have been obtained from simulations 
done with $M=M_0$.}
\begin{ruledtabular}
\begin {tabular}{ccccc} 
 & {\large for all} & & & \\
 &  {\large $[P_a,P_b]$} &
 & {\LARGE
${\delta\!\int\!F_{01}}\over{\left(\delta\!\int\!F_{10}\right)}$ \vline
~${\delta\!\int\!F_{12}}\over{\left(\delta\!\int\!F_{21}\right)}$} & \\
 & {\large within} & & & \\
{\large scope}  & {\large scope of P} & {\Large $\alpha$}{\large $=1$} & 
{\Large $\alpha={1 \over 3}$} & {\Large $\alpha$}{\large $=0$}\\
{\large of $P$} & {\large $P \rightarrow P^*$} & & & \\
\hline
 & & & & \\
{\large $P>P_T$}  & {\large $P \rightarrow P_{b\ominus}$} & 
{\LARGE
$\sim\oslash\over{\left(\oslash\right)}$ 
\vline ~$\oslash\over{\left(\oslash\right)}$} & 
{\LARGE
$\sim\oslash\over{\left(\oslash\right)}$ \vline
~$\oslash\over{\left(\oslash\right)}$} &
{\LARGE
$\sim\oslash\over{\left(\oslash\right)}$ \vline
~$\oslash\over{\left(\oslash\right)}$}\\
 & & & & \\
\hline 
 & & & & \\
{\large $P>P_T$} & {\large $P \rightarrow P_{a\oplus}$} &
{\LARGE
$Si\over{\left(Si\right)}$ \vline ~$\oslash\over{\left(\oslash\right)}$} & 
{\LARGE
$Si\over{\left(\oslash\right)}$ \vline ~$\oslash\over{\left(\oslash\right)}$} &
{\LARGE
$\sim\oslash\over{\left(\oslash\right)}$ \vline
~$\oslash\over{\left(\oslash\right)}$}\\
 & & & & \\
\hline
 & & & & \\
{\large $P<P_T$} & {\large $P \rightarrow P_{b\ominus}$} & 
{\LARGE
$Si\over{\left(Si\right)}$ 
\vline ~$\oslash\over{\left(\oslash\right)}$} & 
{\LARGE
$\sim\oslash\over{\left(\oslash\right)}$ \vline
~$\oslash\over{\left(\oslash\right)}$} &
{\LARGE
$\sim\oslash\over{\left(\sim\oslash\right)}$ \vline
~$\oslash\over{\left(\oslash\right)}$}\\
 & & & & \\
\hline 
 & & & & \\
{\large $P<P_T$} & {\large $P \rightarrow P_{a\oplus}$} &
{\LARGE
$Si\over{\left(Si\right)}$ \vline ~$Si\over{\left(Si\right)}$} & 
{\LARGE
$Si\over{\left(\sim\oslash\right)}$ \vline 
~$\sim\oslash\over{\left(\oslash\right)}$} &
{\LARGE
$Si\over{\left(\sim\oslash\right)}$ \vline
~$\oslash\over{\left(\oslash\right)}$}\\
 & & & & \\
\end {tabular}
\end{ruledtabular}
\end {table*}
\endgroup
Appearance of harmony between the distributions 
$T_0(M) \;,\; T_1(M) \;,\; T_2(M)$ indicates
certain threshold value $M_\tau$  (see Fig.~\ref{fg.10}). 
The REP-ITP simulation series done with $M$ slightly
smaller or slightly greater than $M_\tau\!\approx\!160$ result in
the specifically different increments $\delta \! \int \! F_{01}$.
Similar effect, concerning however the $\delta \! \int \! F_{12}$, 
has been observed for example of the REP-DTP  
series ~(see Table~\ref{tab.2}). 
\begingroup
\squeezetable
\begin{table}
\caption{\label{tab.2}Dependence of increments in MMS on $M$ 
for example of the REP-ITP
with $P=0.1311$, {\large $\alpha$}$={1 \over 3}$, 
(then $M_{\tau}\!=\!160$)
and 
REP-DTP with   
$P=0.104$, {\large $\alpha$}$=1$ (then $M_\tau \approx 270$ ;
see Table~\ref{tab.1} for denotations).}
\begin{ruledtabular}
\begin {tabular}{ccc}
 & & \\
 & {\LARGE
${\delta\!\int\!F_{01}}\over{\left(\delta\!\int\!F_{10}\right)}$ \vline
~${\delta\!\int\!F_{12}}\over{\left(\delta\!\int\!F_{21}\right)}$} & \\
 & &\\
$M<M_{\tau}$ & REP-ITP & $M>M_{\tau}$\\
 & &\\
\hline
 & &\\
{\LARGE
$Si\over{\left(\oslash\right)}$ \vline 
~$\oslash\over{\left(\oslash\right)}$} & &
{\LARGE
$\oslash\over{\left(\oslash\right)}$ \vline
~$\oslash\over{\left(\oslash\right)}$}\\
 & &\\
\hline
 & &\\
$M<M_{\tau}$ & REP-DTP & $M>M_{\tau}$\\
 & &\\
\hline
 & &\\
{\LARGE
$Si\over{\left(\oslash\right)}$ \vline 
~$Si\over{\left(\oslash\right)}$} & &
{\LARGE
$Si\over{\left(\oslash\right)}$ \vline
~$\oslash\over{\left(\oslash\right)}$}\\
 & \\
\end {tabular}
\end{ruledtabular}
\end {table}
\endgroup
Let us recall that in course of REP evolution simulated with 
$M\!\approx \! M_0$, the increments $\delta \! \int \! F_{T_j}$
appear in form of a number of few site clusters and single sites scattered 
throughout whole the $\int \! F$. On the other hand, the corresponding 
spreading percolation SPP (see Section~\ref{sec.4.1}), being the simplest 
process without modeling the finite-size effects which can be compared to the 
REP, results only in dispersed  sedimentation-type increments in the MMS that 
occur only at jumps between steps in the $I_{mms}(T)$ (see Fig.~\ref{fg.10}). 
These increments can be compared to the $\delta \! \int \! F_{T_j}$
that correspond to jumps in the $I_{mms}(T)$ appearing in course of the
REP evolution.  The difference revealed by this comparison
can be attributed  to accomplishment of the bi-state that is peculiar to
modeling the finite-size effects in accordance with the REP covering scheme. 
We have, however, observed that the REP-ITP simulated with the sufficiently 
high $M$ results in the $\delta \! \int \! F_{T_j}$ that, with growing $M$, 
are being reduced to the forms of dispersed sedimentation-type
increments resembling, to some extent, the increments resulting from the SPP 
(see Fig.~\ref{fg.10}). This resemblance does not mean reducing the REP to 
SPP or vanishing the F-SE; the bi-state and its stabilization events at $T_j$ 
are further present. Only effects attributed to {\itshape discreteness} of the 
information transfer have reduced manifestations in the MMS patterns resulting 
from the REP simulations with the sufficiently large $M$. 
This reduction affects diversity of forms of 
increments in the MMS. The form of the $\delta \! \int \! F_{T_j}$ is the one 
of manifestations of the F-SE which is relevant to opportunity for 
employing the REP simulations reported here to facilitate designing methods of 
regulating the collective effect accomplishment in corresponding real systems 
(see Section~\ref{sec.6.3}). For this reason only, vanishing of this 
manifestation with growing $M$ for the REP-ITP simulation series has 
been accepted here as criterion being used to assess upper limit, $M_1$ 
to the values $M$ being moderate.  Accordingly, $260 < M_1 < 280$ has been
identified by appearance of this qualitative change in the increments,  
$\delta \! \int \! F_{T_1}$ and $\delta \! \int \! F_{T_2}$.
The distributions $T_j(M)$ detect here also certain value 
$M^* \approx 220>M_\tau$ such that with, $M^* <M<M_1$, the
$\delta \! \int \! F_{T_j}$ can  be either scattered increments specific to the
moderate $M$ or may resemble the dispersed sedimentation-type increments.
\par
Let us note, eventually, that results of REP-ITP simulations with 
$M\!\approx\!M_1$ reveal manifestations attributed to the discreteness but 
they are remarkably weaker than those obtained with much lower moderate $M 
\geq M_0$ (see Fig.~\ref{fg.10}). Although, the value of $M_1$ depends on  
{\large $\alpha$} and $P$, the experience acquired shows that 
the $M_1$  appears to be sufficiently high for performing reliable simulation 
series of the REP-ITP with varying $P$, {\large $\alpha$} and $M < M_1$.
Occurrence of the $M_1$ has been detected for the REP-ITP and we have not 
observed the specific changes in $\delta \! \int \! F_{T_j}$ for the REP-DTP 
simulated even with very high value M=600. 
Inspection of the REP-DTP and REP-ITP realizations in course of their 
evolutions explains this observation: 
Each realization of the REP-ITP is constituted by chains of islands 
distributed rather uniformly in direction parallel to the initial structure 
(IS), however, some of realizations include narrow elongated tongues 
(see Fig.~\ref{fg.3}). For the very high $M\!\approx\!M_1$, this results 
in mean expected pattern revealing no changes within 
central part of the ~$\int \! F$ at the jump. Then, only the frontal parts of 
the realizations are sufficiently diversified to contribute to the increment 
in the MMS. On the other hand, realizations of the REP-DTP are split 
into two classes: Each realization in the class encompassing of about 
$70\% \div 80\%$ of all the realizations forms only one or two big rugged 
clusters, situated against short subintervals of each IS chain, and few very 
small islands. Realizations constituting the remained part are represented by 
very small areas covered in their domains and resemble less developed 
realizations of the REP-ITP. Because of this diversification, reduction of the 
increment in the MMS at jumps to the form that would resemble this observed 
for the REP-ITP simulated with $M \! \approx \! M_1$, would require values of 
$M$ much higher than the $M_1$ detected for the REP-ITP.
\subsection {\label{sec.6.3}Discussion and conclusions}
Series of the MMS patterns $\int \! F(T)$ represents accomplishing the model
collective effect in course of the REP.
Discreteness of the REP results in a sequence of jumps in the MMS development
during the process evolution. Modeling of the finite-size effects results
in appearance of the structural bi-state. Using structural characteristics of
the MMS pattern enables us to identify stabilization of the bi-state at
the stage denoted as $T_2$ which corresponds to certain jump
~(see Sections~\ref{sec.5.2}, \ref{sec.5.3}).
This $T_2$ is the reference stage used in the search for specific changes in 
features of the MMS evolution pattern
with varying the process parameters $\{P$, {\large $\alpha$}, $M\}$.  
The $\tau_{01}$ and $\tau_{12}$ separate the triplet of subsequent neighbor 
jumps in values of the MMS characteristics which occur in course
of the MMS development at stages $T_j \in \{T_0$, $T_1$, $T_2\}$ 
~(see Fig.~\ref{fg.4}). 
Occurrence of nonnull sedimentation-type increments and
reductions in the MMS for the accumulation periods $\tau_{01}$ and $\tau_{12}$ 
is specific feature attributed to modeling the finite-size effects in the REP
~(see Fig.~\ref{fg.6} and Tables~\ref{tab.1},~\ref{tab.2}).
Specific form of the scattered increments $\delta \! \int \! F_{T_j}$
in the MMS development, which correspond to the jumps, is characteristic 
manifestation of the finite-size effects modeling in the REP.
Reduction of this particular manifestation of the finite-size effects in a 
series of the REP-ITP simulations due to substantial reduction of the 
discreteness manifestations in the MMS patterns, which is observed
with the sufficiently large $M$, corresponds to the 
upper bound $M_1$ of the $M$ being moderate. The lower bound $M_0$ assures 
that all the REP realizations have no covered site in common and features of 
the corresponding MMS patterns are reliable.
\par
Accordingly, the MMS evolution pattern $\int \! F(T)$, computed 
with the $M$ being moderate $M_0 \!\leq\! M\!<\! M_1$,
appears as sequence of bursts $\delta \! \int \! F_{Tj}$ in series of patterns 
of the MMS increments. The patterns representing these bursts separated by the 
accumulation periods are constituted of sites scattered throughout whole
the $\int \! F(T_j)$. 
This specific form of the increment $\delta \! \int \! F_{Tj}$ enables us 
distinguishing the $\delta \! \int \! F_{Tj}$ from neighbor increments for 
the accumulation periods by performing only observation of the MMS increment 
patterns (see Fig.~\ref{fg.6}).
\par
In certain conditions parameterized by, $\{P$, {\large $\alpha$}, $M\}$, a 
burst $\delta \! \int \! F_{T_j}$ separating an accumulation period with the
sedimentation-type increment in the $\int \! F$ from the following one
with no or vanishing increment in the $\int \! F$ can detect the $T_1$.
Here, the examples of such conditions have been found by performing a number 
of series of the REP simulations ~(see Tables~\ref{tab.1},~\ref{tab.2}). 
These results suggest that computing a number of series of the REP 
representing adequately a process underlying evolution of certain real system 
ensemble would enable us to predict parameters of 
real conditions corresponding to values of the relevant model parameters,  
$\{P$, {\large $\alpha$}, $M\}$. In those real conditions,
stabilization of the structural bi-state of the $\int \! F(T)$ 
recorded in course of the real process would be identified by observing the
sequence of the bursts  $\delta \! \int \! F_{T_j}$ and increments in the MMS
which may occur for periods between them. Note in this connection that 
identification of the increment in the MMS at a jump $T_j$ or for the 
accumulation period refers to the same state of the bi-state that can be 
easily found in the simulation results. In a real system, performing this 
task may require more effort, particularly in situation when differences 
between states of the bi-state, $\Delta \! \int \! F(T,T+1)$ 
are comparable to the increment in MMS at the jump, 
$\delta \! \int \! F_{T_j}$. However, for the jumps between the stabilized 
bi-states, the covered area representing the difference,  
$\Delta \! \int \! F(T,T+1)$  is distanced from the front region of the 
expanding MMS whereas the increment, $\delta \! \int \! F_{T_j}$ consists of 
elements occurring within whole the MMS (see Fig.~\ref{fg.6} and 
Section~\ref{sec.5.3.3}). This difference between the sets noted is not 
observed for the pure ITP process, REP-ITP-P, in which stabilization of the 
bi-state does not take place. Eventually, the feature just recalled would help 
detecting the $T_2$ in real systems corresponding to the REP-ITP or REP-DTP. 
However, the distributions of the $T_j(P)$ with other parameters fixed are 
much more regular for the REP-ITP than for the REP-DTP with lower $M$ 
(see Fig.~\ref{fg.8}) and simpler with higher $M$. This makes real processes 
with scheme of information transfer corresponding to the REP-ITP the first 
candidates for the regulation. Thus results of the computations would enable 
us reducing a number of experiments required to realize the real process with 
regulating the collective effect accomplishment.
\par
Let us emphasize that the opportunity just revealed appears for processes with 
accomplishing the bi-state enabling us detecting its stabilization events 
$T_j$ whose role has been elucidated here previously. We have also shown that 
realization of transferring the information portions by the 
hopping finite-size elements is a prerequisite for appearance of the bi-state.
Thus the opportunity aforementioned will not appear for processes without this 
kind of the information transfer.
\par 
Eventually, we {\itshape conclude}: 
Pattern of the MMS development, which represents accomplishing
the model collective effect in course of the REP, experiences predictable 
changes with varying the parameters $(P,\; \alpha,\; M)$ specifying 
the model ambient conditions.
\par 
This feature is attributed to modeling the finite-size effects
in the REP simulation series. 
\par
For the REP-ITP, the effective prediction is 
possible with moderate number of realizations, $M_0 \!\leq\! M\!<\! M_1$.
\subsection {\label{sec.6.4}Final remark}
Let us recall that manifestations of the finite size effets (F-SE) appear to 
be a prerequisite for using changes in ambient conditions to regulate 
accomplishing the collective effect resulting from evolution of the ensemble 
of discrete systems. The F-SE result from fact that the local information 
transfer is realized with participation of elements represented by pairs of 
zero dimensional entities identified at the nearest sites of the array and 
experiencing conditional hopping to positions close by while transferring 
certain information portions and preserving their orientation in the space. 
Diversity of opportunities for realizing information transfers by the hopping 
pairs, which is much higher than for only distinct hopping zero dimensional 
particles, required effective modeling whose spatiotemporal aspect has been 
explained in Section~\ref{sec.2}. This has enabled us to reveal the way of 
regulating the collective effect accomplishment. 
\par
On the other hand, it is known that certain evolutions of model discrete 
systems, whose all elements participating in the information transfer process 
allow considering them as zero dimensional particles only, may result in 
self-organizing structures composed of sites indicated in certain way in 
course of the process (see eg. Ref.~\cite{8}). However, the appearing 
structures do not start acting as the hopping {\itshape functional} elements 
if no additional assumptions concerning that functioning in further course of 
the model process are made. This seems to suggest that the pairs representing 
the functional elements are required from the beginning of evolution of the 
ensemble systems to afford the opportunity for employing the ambient 
conditions to regulate the emerging collective effect in the way reported here.
This particularly concerns the task of investigating the influence which 
changes in ambient determined factors can have on collective outcome of 
complex processes taking place in systems of the ensemble. The possible 
evolution of swarms of nanoelements mentioned here (see Sections 
~\ref{sec.1},~\ref{sec.2}) as well as turbulent transport reported previously 
by us \cite{9} may be considered as the examples of real processes for which 
this opportunity for investigating them appears to be useful. 

\appendix*
\section {\label{ap}Mean Measure Set Representation}
Below, we present only brief technical way, in which the mean measure set
representation, MMS is computed; the relevant theory is given in \cite {11}.
\par
The MMS , denoted here also as $\int \! F$
has been proposed  \cite {11} for characterizing mean expected form of a finite
random set $F=\{A_i, ~i=1,2,...,M\}$ constituted by finite number $M$ of its 
realizations $A_i$, 
\begin{equation}
\int \! F=\{\Phi_M > h\}^1 \cap \{\Phi_M < h\}^0 , \label{eq.a1}
\end{equation}
with the value $h \in [0,1]$ given by the inequalities,
\begin{equation}
\mu \{\Phi_M > h\} \leq {1\over M} \mathop {\sum_{i=1}^{M}} \mu \{A_i\} \leq
\mu \{\Phi_M \geq h\}, \label{eq.a2}
\end{equation}
where $\in$ is read "is an element of" , $\mu \{.\}$ denotes the number of 
elements of
the set $\{.\}$,  $\{\Phi_M > h\}$ is a set of such elements $x \in \chi$ that
$\Phi_M(x) > h$ and symbols $\{.\}^0$, $\{.\}^1$  denote set-theoretical cells
determined in respect to a set $\{.\}$.
A cell of empty conjunction, 
$D^0=\{\mathcal {A} \subset \mathcal {M} ~:~ \mathcal {A} \cap D= \oslash \}$
and the cell of inclusion,
$D^1=\{\mathcal {A} \subset \mathcal {M} ~:~ \mathcal {A}^c \cap D= \oslash \}$
are defined as classes of finite sets $\mathcal {A}$ in the space 
$\mathcal {M}=2^\chi$  of all subsets of the space $\chi$, in respect to 
elements D
of the set $\mathcal {D}$ covering the space $\chi$ where, 
$\mathcal {D}=\{D_j \subset \chi , ~ j=1,2,...,N_{\mathcal {D}}\}$ and  
$D_k \cap D_j = \oslash$ ~$\forall j \neq k$.
Here, $\mathcal {A}^c$ denotes complement of the $\mathcal {A}$ to $\mathcal 
{M}$,
symbol $\oslash$ denotes a null set, $\forall$ is read 
"for all", $\cap$ and $\subset$ denote common part and inclusion of sets,
respectively. 
The $\Phi_M=\Phi_M(x)$ is the mathematical expectation,
\begin{equation}
\Phi_M(x)= {1\over M} \mathop {\sum_{i=1}^{M}} \Phi(x,A_i) , \label{eq.a3}
\end{equation}
with
$\Phi(x,A_i)=0 ~if~ A_i \in \{x\}^0$
and 
$\Phi(x,A_i)=1 ~if~ A_i \in \{x\}^1$,
where $\{x\}$ is the single element set. Uniqueness of the canonical cell
representation Eqs.~(\ref{eq.a1}) has been proved \cite {11}.
The multitude of sets $\int \! F$ is represented unambiguously by
its single component: if $h \geq 0.5$ one takes the one whose measure is 
largest;
if $h < 0.5$ one takes the one whose measure is smallest \cite {11}. 
In this work, only such single set representation is used so it is denoted in 
the
text by $\int \! F$ or MMS. 

\begin{acknowledgments}
The parallel computations have been performed by using 
computer Cray-T3E with computational grant at The Interdisciplinary Centre for
Mathematical and Computational Modelling at The Warsaw University; Partial 
support
from Polish-British Research Partnership Programme WAR/341/202 and
NATO Collaborative Linkage Grant PST.CLG.976545 is acknowledged.
\end{acknowledgments}

\bibliography{W-Kozlowski-arxiv}

\end{document}